\documentclass[useAMS,usenatbib, twocolumn]{mn2e}
\usepackage[pdftex]{graphicx}
\usepackage{txfonts}
\usepackage{bm}
\usepackage{subfigure}
\usepackage{multicol}
\usepackage[normalem]{ulem}

\def\flux{\rm erg~s$^{-1}$~cm$^{-2}$}

\def\rate{\rm g~s$^{-1}$}
\def\vel{\rm cm~s$^{-1}$}

\def\beq#1{\begin{equation}\label{#1}}
\def\eeq{\end{equation}}
\def\beqa#1{\begin{eqnarray}\label{#1}}
\def\eeqa{\end{eqnarray}}

\def\Eq#1{Eq.~(\ref{#1})}
\def\Fig#1{Fig.~\ref{#1}}
\def\eqn#1{~(\ref{#1})}
\def\myfrac#1#2{\left(\frac{#1}{#2}\right)}

\def\comment#1{\relax}

\title[Continuum in X-ray pulsars]{On the dependence of the
X-ray continuum variations with luminosity in accreting X-ray pulsars}

\author[K. Postnov et al.] {K. A. Postnov$^{1,4}$,
M. I. Gornostaev$^{1,4}$, D. Klochkov$^2$, E. Laplace$^2$, V. V. Lukin$^{3}$, N. I. Shakura$^4$
\thanks{E-mail: pk@sai.msu.ru }\\
$^{1}$ Faculty of Physics,
M. V. Lomonosov Moscow State University,
Leninskie Gory, Moscow 119991
Russia\\
$^{2}$ Institute of Astronomy and Astrophysics, Karl-Eberhard University, T\"ubingen, Sand 1, D-72076, Germany\\
$^{3}$ M.V. Keldysh Institute of Applied Mathematics RAS, Miusskaya sq., 4, Moscow,
Russia\\
$^{4}$ Sternberg Astronomical Institute, Moscow M.V. Lomonosov State University, Universitetskij pr., 13, 119992, Moscow, Russia\\
}

\begin{document}

\date{Received ... Accepted ...}
\pagerange{\pageref{firstpage}--\pageref{lastpage}} \pubyear{2012}

\maketitle

\label{firstpage}

\begin{abstract}
Using RXTE/ASM archival data, we investigate the
behaviour of the spectral hardness ratio as a
function of X-ray luminosity in a sample of six transient X-ray
pulsars (EXO 2030+375, GX 304-1, 4U 0115+63, V 0332+63, A 0535+26
and  MXB 0656-072).
In all sources we find that the spectral hardness ratio
defined as $F_{5-12\mathrm{keV}}/
F_{1.33-3\mathrm{keV}}$
increases with the ASM flux (1.33--12 keV) at low luminosities and then
saturates or even slightly decreases above some critical X-ray 
luminosity falling into the range $\sim(3-7)\times10^{37}$~erg~s$^{-1}$. 
Two-dimensional
structure of accretion columns in the radiation-diffusion limit
is calculated for two possible geometries (filled and hollow cylinder)
for mass accretion rates $\dot M$ ranging from $10^{17}$ to 
1.2$\times 10^{18}$~g s$^{-1}$.
The observed spectral
behaviour in the transient X-ray pulsars 
with increasing $\dot M$ can be reproduced by
a Compton saturated
sidewall emission from optically thick magnetized accretion columns
with taking into account the emission reflected from
the neutron star atmosphere. At $\dot M$ above some critical value 
$\dot M_{cr}\sim (6-8)\times 10^{17}$~g~s$^{-1}$,
the hight of the column becomes such that the contribution of the reflected 
component to the total emission starts decreasing, which leads to the 
saturation and even slight decrease of the spectral hardness.   
Hollow-cylinder
columns have a smaller height than the filled-cylinder ones, and 
the contribution of the reflected component  
in the total emission does not virtually change with $\dot M$ 
(and hence the hardness of the continuum monotonically 
increases) up to higher mass accretion rates than $\dot M_{cr}$ 
for the filled columns.

\end{abstract}


\section{Introduction}
\label{intro}

Accretion of matter onto a gravitating centre is a ubiquitous physical process in astrophysics. The matter gravitationally captured from the surrounding rarefied medium by a body of mass $M$ (or, if in a binary system, escaping from the secondary star as a stellar wind or due to Roche lobe overflow) eventually reaches the surface of the body with the velocity of the order of the parabolic one, $v=\sqrt{2GM/R}$, where $R$ is the radius of the body. This velocity becomes very high in the case of accretion onto a compact star.
%(white dwarf, $R_{WD}\sim 1/100 R_\odot$, neutron star, $R_{NS}\sim 10~\mbox{km}$, or %black hole $R=R_g=2GM/c^2\sim 3\hbox{km} (M/M_\odot)$).
For neutron stars with radius $R_{NS}\sim 10~\mbox{km}$ the velocity reaches $\sim 10^{10}~\mbox{\vel}$, so that the accretion energy release
is $L\sim 0.1 \dot Mc^2$, where $\dot M$ is the mass accretion rate and $c$ is the velocity of light. These basic facts required the calculation of the accreting matter braking near the compact star surface. In the pioneer work by
\cite{1969SvA....13..175Z}, the structure of the shock wave produced by
spherically symmetric accreting matter near the neutron star surface without magnetic field was calculated. Braking of the accreting matter was assumed to be primarily due to Coulomb interactions of the infalling protons with the surface, which is justified at low accretion rates when interaction of the charged particles with radiation is insignificant.

%when the accretion rate is much smaller than the Eddington rate $\dot M_c\simeq 10^
%{18}~\mbox{\rate}$.

\begin{figure*}
\center{\includegraphics[width=0.9\textwidth]{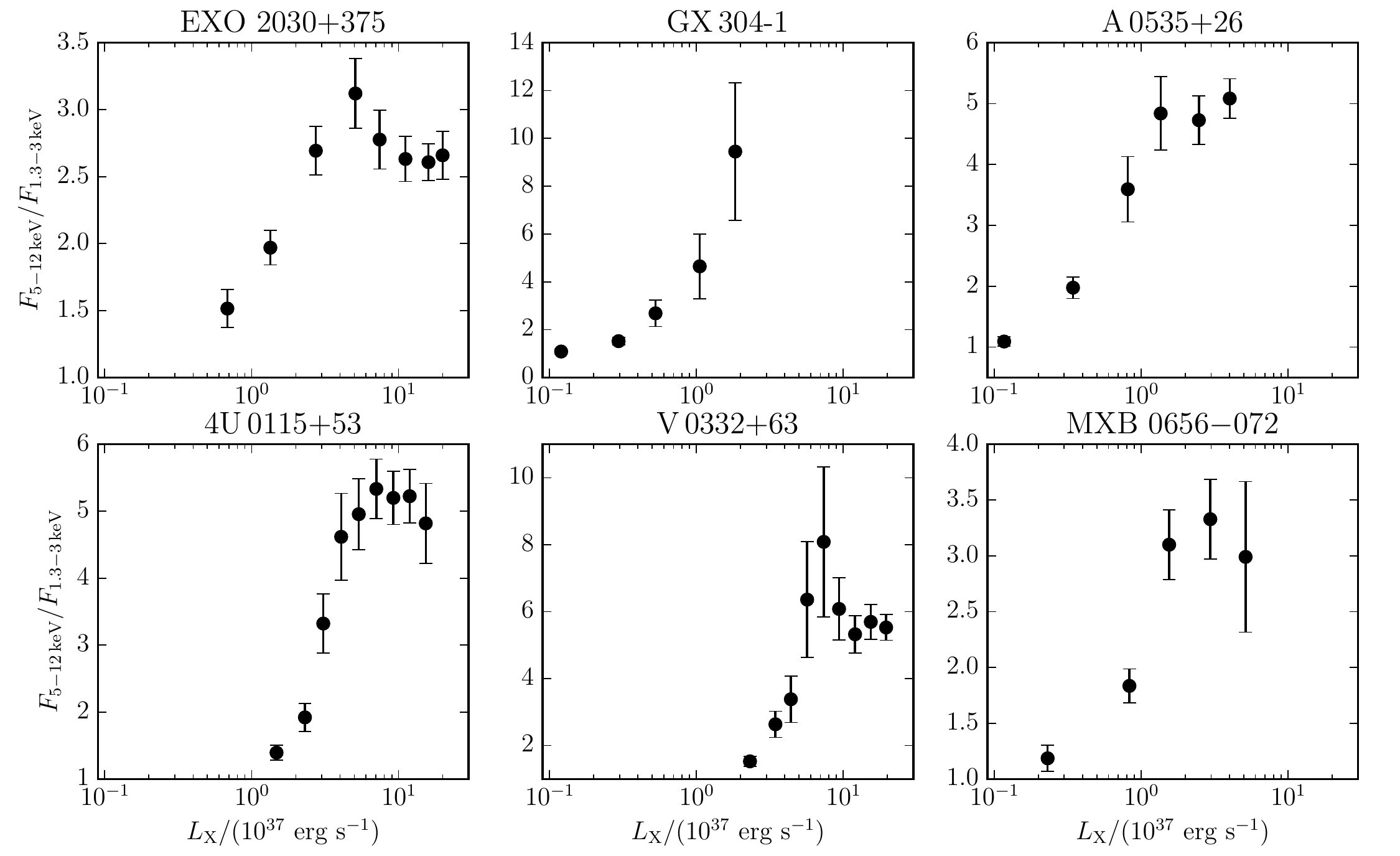}}
%article_xlog_2.pdf}}
\caption{The ratio of the fluxes in 5--12\,keV and 1.33--3\,keV ranges
  (referred to as ``hardness ratio'' in text) measured with RXTE/ASM for
  six transient accreting pulsars as a function of the total ASM flux in the
  1.33--12\,keV range.
}
\label{f:hardness}
\end{figure*}

Immediately after the discovery of X-ray pulsars in 1971 \citep{1971ApJ...167L..67G} it was realized that in bright pulsars the radiation plays a crucial role in braking of the accreting matter onto the surface of a neutron star with strong magnetic field. Indeed, in this case the matter is canalized by the neutron star magnetic field onto the magnetic polar caps, which are characterized by a small radius $r_0\approx R_{NS}\sqrt{R_{NS}/R_A}\sim 10^5$~cm (for a purely dipole field), where $R_A$
%\sim 10^{8}\hbox{cm}\mu_{30}^{4/7}\dot M_{17}^{-2/7}$ 
is the magnetosphere radius. 

It is now well recognized that once the radiation density above the polar cap
starts playing the role in the accreting matter dynamics, an optically thick
accretion column above the polar cap is formed \citep{BaskoSunyaev76} (see \cite{2015MNRAS.447.1847M}
for the latest investigation). The characteristic height of the column increases with accretion rate,
most of the emission escapes through sidewalls,
and it can be expected that above some X-ray luminosity the dependence of the observed properties
of the continuum emission on the X-ray luminosity can become different from
that in the low-luminosity regime.

Here we stress that the effects of the strong magnetic field 
are important for the observed properties of X-ray continuum at energies 
below the cyclotron line energy due to 
scattering processes which have different cross-sections for different photon
polarization modes. 
At small accretion rates, the accretion mound can be viewed as a thin slab, and 
mostly ordinary photons (for which the electron scattering cross-section 
is strongly angular dependent but frequency independent) escape along the magnetic field lines forming 
a pencil-beam X-ray emission diagram \citep{1975A&A....42..311B}. In contrast, when 
the accretion column is formed at higher X-ray luminosities, mostly extraordinary photons 
(for which the electron scattering cross-section is angle independent but strongly frequency dependent) 
escape sidewall to form a fan-like X-ray emission beam \citep{1986Ap.....25..577L}.

Presently, there is a wealth of observational data on X-ray pulsars
that can be used to check different regimes of accretion onto
magnetized neutron stars.
Different types of spectrum--luminosity dependence were first
discovered based on the behavior of cyclotron resonant scattering
features (CRSFs) (e.g. \cite{2006MNRAS.371...19T, 2007A&A...465L..25S, Klochkov:etal:11, Becker_ea12, 2014ApJ...781...30N}).

However, the mechanism of the CRSF generation in accretion columns
of accreting neutron stars is still not well understood due
to a very complicated physics of radiation transfer in super-strong
magnetic fields of neutron stars.
For example, recently a novel model for CRSF production
in spectra of bright X-ray pulsars due to
reflection of the radiation generated in the accretion column
from the neutron star surface was proposed \citep{2013ApJ...777..115P}. This model
can successively explain the CRSF energy dependence on luminosity observed
in outbursts of the transient X-ray pulsar V~0332+53 (see also \citep{2015MNRAS.448.2175L} for further
applications of the model).
From the observational
point of view, the study of CRSFs is a challenging task because it requires long
dedicated observations using the instruments with high effective area
and spectral resolution above 20--30\,keV. Only in a fraction of
accreting pulsars CRSFs are reliably measured \citep{2012MmSAI..83..230C, 2014SSRv..tmp...58R}.

Recently, \cite{Klochkov:etal:11} and \cite{Reig:Nespoli:13}
pointed out that
in addition to the CRSF behaviour,
different types of the spectrum--luminosity dependence
in accreting pulsars can be probed by studying the variability of the X-ray
continuum with luminosity.
Specifically,
it was found that the X-ray continuum
described by a power-law dependence $F_\mathrm{X}\sim E^\Gamma$ becomes \textit{harder} with
increasing X-ray flux in X-ray pulsars showing a \textit{positive} correlation of
the CRSF energy $E_c$ with flux,
whereas it becomes \textit{softer} or virtually does not change with increasing flux
in pulsars showing a negative $E_c(F_\mathrm{X})$-correlation
(Figs.\,3 and 6 in \cite{Klochkov:etal:11}). Clearly, the continuum studies
are much less demanding from the point of view of data analysis than the spectroscopic
analysis of the CRSF behaviour (if the CRSF is reliably identified in the
X-ray pulsar spectrum at all).

%In this work, we concentrate on the continuum
%dependence on luminosity in accreting pulsars.
%}

The purpose of the present paper is to perform model 2D calculations of
the accretion column structure in the radiation diffusion approximation
and to calculate the emerging X-ray continuum
in order to see how
the hardness of the sidewall emission
from accretion columns with different geometries (a
filled or hollow cylinder) changes with
accretion rate variations in the regime of radiation-dominated
accretion flow braking. We calculate the sidewall energy flux escaping 
from the optically thick accretion column and the saturated Compton  
spectrum of extraordinary photons. We find that 
for typical parameters of X-ray pulsars (the neutron star surface 
magnetic field $B= 3\times 10^{12}$~G, mass accretion rate onto one pole 
$\dot M=(1-12)\times 10^{17}$~g~s$^{-1}$), with increasing mass accretion rate the hardness of 
the emergent sidewall emission from the column always increases. 
The sidewall emission from the column is Doppler-boosted by unbraked 
matter in the external part of the column toward the neutron star surface. 
The reflected radiation 
from the neutron star atmosphere calculated in the single-scattering approximation
is harder than the incident emission, and its addition to the 
total flux form the column renders the spectrum even harder. 
However, starting from some accretion rate the height of the column
increases such that the fraction of the reflected radiation in the
total flux starts decreasing. This purely geometrical effect  
leads to the saturation of the hardness ratio 
at $\dot M_{cr}\sim (6-8)\times 10^{17}$~g~s$^{-1}$ (for the assumed neutron star
parameters).  
In the case of the hollow-cylinder accretion column geometry, the saturation 
of the spectral hardness has not been reached in the calculated range of
mass accretion rates due to smaller photon diffusion times across the 
columns thickness and correspondingly smaller heights of such columns, when
the contribution of the emission reflected from the neutron star surface 
always dominates.  

The structure
of the paper is as follows.
%In Section \ref{s:crlum} we remind the
%definition of critical luminosities separating local %sub-Eddington and
%super-Eddington regimes.
In Section \ref{s:obs} we briefly describe
the observed correlations of CRSF with X-ray flux and X-ray continuum
properties with X-ray flux. In Section \ref{s:geom} the basic
equations of the structure of 
the radiation-supported columns 
and their numerical solution are described. The results of
calculations of the spectrum are presented in Section \ref{s:spectrum}, which is followed by the
summary and conclusions.

\section{Observations}
\label{s:obs}

There is a growing number of X-ray pulsars showing
the dependence of $E_c$ of their X-ray flux (e.g., \cite{Becker_ea12}
and references therein).
The cyclotron energy $E_c$ in the bright transient X-ray pulsar
V\,0332+53 decreases with luminosity $L_{\rm X}$
\citep{2006MNRAS.371...19T, 2010MNRAS.401.1628T}.
A similar negative $E_c(L_{\rm X})$-correlation has been reported for
another bright transient pulsar 4U\,0115+63 \citep{2007AstL...33..368T, 2004ApJ...610..390M}.
Recently, however, this correlation has been questioned by
\cite{2013A&A...551A...6M}, who argued
that the $E_c(L_{\rm X})$-correlation in 4U\,0115+63 is an artefact
related to the continuum modelling (see also \cite{2013AstL...39..375B}
for an independent analysis).

There is another type of X-ray pulsars with lower X-ray luminosities
and somewhat higher magnetic
fields which exhibit a \emph{positive} correlation of $E_c$ with $L_{\rm X}$:
Her X-1 \citep{2007A&A...465L..25S}, GX\,304$-$1 \citep{2011PASJ...63S.751Y, 2012A&A...542L..28K},
and probably Vela X$-$1 \citep{2014ApJ...780..133F}
and A\,0535+26 \citep{Klochkov:etal:11, 2013A&A...551A...6M}.

As mentioned in the Introduction, a similar 
bimodality is observed in the dependence 
of the X-ray continuum hardness on luminosity. This dependence can be studied using, 
for example, data from all-sky monitors such as RXTE/ASM and MAXI, i.e. without dedicated
spectroscopic observations which are necessary to measure
CRSFs. Following this approach, we measured the hardness ratio as a
function of luminosity in the accreting
pulsars GX\,304$-$1, 4U\,0115+63, V\,0332+53, EXO\,2030+375, A\,0535+26
and  MXB\,0656−072
using the data from different energy bands of RXTE/ASM. The result
is shown in Fig.\,\ref{f:hardness}.  To convert the ASM count rates into X-ray
luminosities, we used published distances to the sources
(GX\,304$-$1: $\sim$2\,kpc, \citealt{Parkes:etal:80}; 4U\,0115+63: $\sim$7\,kpc,
 \citealt{Negueruela:Okazaki:01}, V\,0332+53: $\sim$7\,kpc,
  \citealt{Negueruela:etal:99}; EXO\,2030+375: $\sim$7\,kpc,
  \citealt{Wilson:etal:02}; A\,0535+26: $\sim 2$~kpc, 
\citealt{1998MNRAS.297L...5S}; MXB\,0656-072: $\sim 3.9$~kpc,
\citealt{2006A&A...451..267M})
  and their broadband X-ray spectra from the archival pointed
  observations with INTEGRAL and RXTE available to us.
One can see that at lower fluxes
  the hardness ratio increases with flux. At a certain flux, however, a
  flattening of the hardness ratio is observed in
  4U\,0115+63, V\,0332+53, EXO\,2030+375, A\,0535+26
and  MXB\,0656−072. This `turnover'
  occurs at the flux roughly corresponding luminosities $(3-7)\times 10^{37}$\,erg s$^{-1}$. In GX\,304$-$1,
  which does not show such a turnover, this
  `critical' luminosity
  has simply not been reached during the outbursts registered by RXTE ASM.

A similar dependence of spectral properties on the X-ray flux at luminosities below 
$\sim 10^{37}$~erg~s$^{-1}$ was also reported from spectroscopic observations 
of the transient Be-X-ray pulsar GRO J1008-57 \cite{2014EPJWC..6406003K}.

At the lowest energies covered by RXTE/ASM, the spectrum of accreting
pulsars is significantly affected by photo-electric absorption caused
mostly by matter in the vicinity of the source. Such an absorption
naturally influences the derived hardness ratios and cannot be
accounted for without a high-resolution spectral analysis using pointed
observations. Neutral absorbing matter surrounding the
accretor might correlate with the mass transfer rate in the binary and
might therefore be proportional to the X-ray luminosity of the pulsar.
Such an effect might at least contribute to the observed positive correlation
between the hardness ratio and flux. However, we do not expect any
break in the slope of the correlation between the absorption column
density and flux at $L_{\rm X}\sim 10^{37}$~erg s$^{-1}$ (although we cannot
exclude this possibility). We therefore identify the observed breaks
with those reported by \cite{Reig:Nespoli:13} at similar X-ray luminosities
based on the data from RXTE/PCA observations taken at
higher energies which should be unaffected by the photo-electric absorption.

\section{Accretion column structure in the radiation diffusion limit}
\label{s:geom}

\subsection{Qualitative considerations}

At high luminosities, $L_{\rm X}\gtrsim 10^{37}$~erg~s$^{-1}$, the radiation pressure dominates, 
and the braking of the accreting matter flow is due to interaction with photons. In this case,
the kinetic energy of falling particles $m_pv_0^2/2$, where $v_0=10^{10}$~cm s$^{-1}$ is the free-fall velocity near the NS surface, decreases by almost 99\% at optical depths
$\tau\sim (2-3)(c/v_0)\sim 4-9$. This was shown for the first time by \cite{1973NPhS..246....1D},
and in a wide range of parameters was calculated by \cite{1981A&A....93..255W}
by solving two-dimensional
gas-dynamic equations with diffusive radiation transfer.   As a result, the shape of
the zone of energy release looks like a 'mound' with the characteristic height $z_0\sim r_0$.
%unlike in the Coulomb braking case. 

With further increasing X-ray luminosity above $L_{\rm X}>L^*$,
as was first shown by \cite{BaskoSunyaev76},
the height of the braking zone starts increasing, and instead of
a 'mound' an optically thick accretion column appears with $z_0>r_0$.
Most of the
kinetic energy of the accreting flow is transformed into heat
at around $z_0$ (which sometimes is referred to as the
height of the radiation-dominated shock), and below this height the matter settles down
towards the column base. The thermal energy is advected by the settling matter with
an effective velocity of $v\sim 0.1 v_0$, and the generated heat
finally escapes through the column sidewalls.

It is possible to estimate the critical luminosity $L^*$ by equating the
dynamical time of the fall of matter 
to the diffusion time of photons across the column.
The characteristic settling time is
clearly
$t_s=z_0/v$. The characteristic time of the radiation diffusion across the 
filled cylinder column is
\beq{tdif}
t_d\simeq \frac{r_0^2}{\nu_d}\simeq \frac{r_0^2\varkappa_\perp}{c}\rho=\frac{r_0^2\varkappa_\perp}{c}\frac{S}{v}
\eeq
where we have eliminated the density $\rho$ using the mass continuity equation
\beq{e:cont}
S=\rho v=\frac{\dot M}{\piup r_0^2}\,.
\eeq
In the case of a hollow cylinder with wall thickness $br_0$ the diffusion time decreases 
with $b$:
\beq{tdif_hollow}
t_d\simeq \frac{b^2r_0^2}{\nu_d}\simeq \frac{b^2r_0^2\varkappa_\perp}{c}\rho=\frac{b^2r_0^2\varkappa_\perp}{c}\frac{S}{v}=\frac{b\varkappa_\perp}{c}\frac{\dot M}{\piup v}
\eeq
By equating $t_s=t_d$ and noting that the velocity is crossed out, we arrive at
the expression
\beq{}
z_{0,f}=\frac{\varkappa_\perp}{\piup c}\dot M\,,\quad z_{0,h}=\frac{b\varkappa_\perp}{\piup c}\dot M\,
\eeq
for the filled and hollow cylinder geometries, respectively.

Now, by equating $z_0=r_0$, we find the critical mass accretion rate $\dot M^*$
and the corresponding X-ray luminosity
\beq{L*}
L^*=0.1 \dot M^* c^2=
\frac{\piup c}{b\varkappa_\perp}0.1 c^2 r_0\approx 2.36\times 10^{36}[\mathrm{erg\, s}^{-1}]
\myfrac{r_0}{10^5\hbox{cm}}\myfrac{b\varkappa_\perp}{\varkappa_T}^{-1}.
\eeq
Here the factor $b$ should be set one for the filled cylinder geometry. 

It is important to realize that
the accurate value of $L^*$ should be found from two-dimensional numerical
calculations of the radiation transfer equations, and, for example, \cite{1981A&A....93..255W} found
\beq{L*wang}
L^*\approx 5.25\times 10^{36}[\mathrm{erg\, s}^{-1}]\myfrac{E_c}{\bar E}^{2/5}\myfrac{r_0}{10^5\hbox{cm}}\myfrac{M}{1.5 M_\odot}
\eeq
Here the factor $(E_c/\bar E)$ takes into account the effective change in the
photon scattering cross-section in the strong magnetic field and the deviation of the magnetic 
field line geometry from the simple cylindrical case near the NS surface. This factor is $\sim 1$ if
the mean photon energy $\bar E>E_c$. It is seen that our simple estimate \Eq{L*} derived
above is consistent with this value to within a factor of two.
Recently, the critical luminosity was reassessed by \cite{2015MNRAS.447.1847M}
who found it to vary by almost an order of magnitude (depending on the magnetic field)
within an interval centred at
$\sim 10^{37}$~ erg~s$^{-1}$.

Therefore, the qualitative considerations presented above suggest the increase 
in the accretion structure height $z_0\propto \dot M$ above some critical 
luminosity $\sim 10^{37}$~ erg s$^{-1}$ 
(see also a more detailed derivation in \cite{1992ApJ...388..561A}, Section 3.). 

\subsection{Numerical simulations of axially symmetric accretion column}

Qualitative considerations described above can be made more precise by numerical calculations of steady-state accretion columns. This radiation-hydrodynamic problem, complicated by the need to calculate the radiation transfer in a strong magnetic field, has not been solved self-consistently as yet. However, the structure of an optically thick accretion column at luminosities above $10^{37}$~erg~s$^{-1}$ can be calculated in the radiation diffusion approximation with grey scattering coefficients along and perpendicular to the magnetic field lines \citep{1973NPhS..246....1D, BaskoSunyaev76, 1981A&A....93..255W}. The emergent radiation spectrum will be formed due to scattering of photons on electrons in the optically thin outer layers, which is a separate problem (see the next Section).

We shall consider two possible axially symmetric geometries of accretion columns: a filled cylinder of radius $r_0$
and a hollow cylinder of the inner radius $r_0$ and thickness $br_0$.
In both cases $r_0$ is determined by the magnetosphere radius $R_A$, and in the second case the wall thickness $br_0$ can be related to the distance the disc matter enters the magnetosphere before being completely frozen into the magnetic field. We shall assume a purely dipole magnetic field and the disc matter frozen at a fixed fraction of the Alfv\'en radius.

From the definition of the Alfv\'en radius, $R_A \propto {\dot M}^{-\frac{2}{7}}$, and the expression for the polar cap radius $r_0$ we see that the polar cap radius scales with accretion rate as $r_0 \propto {\dot M}^{\frac{1}{7}}$.
The freezing depth of disc plasma at the Alfv\'en radius is fixed as $\Delta = 0.1 R_A$, and since $\frac{\Delta}{R_A} = b$ we have $b \simeq  0.1$. This means that with changing mass accretion rate the thickness of the accretion wall in the case of hollow cylinder geometry changes as
$br_0\propto{\dot M}^{\frac{1}{7}}$ \footnote{In principle, one may consider further complications, e.g., using a specific model of the magnetosphere-disc coupling, like it was done in \cite{2015MNRAS.447.1847M}. This only slightly changes the dependence of $br_0$ on $\dot M$.}.

Note that the value of $b$ is difficult to calculate precisely; it is clear that it should depend on many factors, including the
NS magnetic field axis misalignment, the details of the matter
entering the magnetosphere due to various instabilities, etc., which are beyond the
scope of the present study. However, we should note that in the case of thin cylinder wall, $ b\ll 1$, the radiation diffusion time \Eq{tdif} is proportional $t_d\propto b$, thus decreasing the height of the column $z_0$ at a given mass accretion rate $\dot M$. This effect is confirmed by our calculations (see below).
The neutron star magnetic field is assumed to be homogeneous across the column. This approximation is justified for not very high accretion heights (especially in the case of the hollow cylinder geometry).

\begin{figure*}
\begin{multicols}{2}
\includegraphics[height=0.4\textwidth]{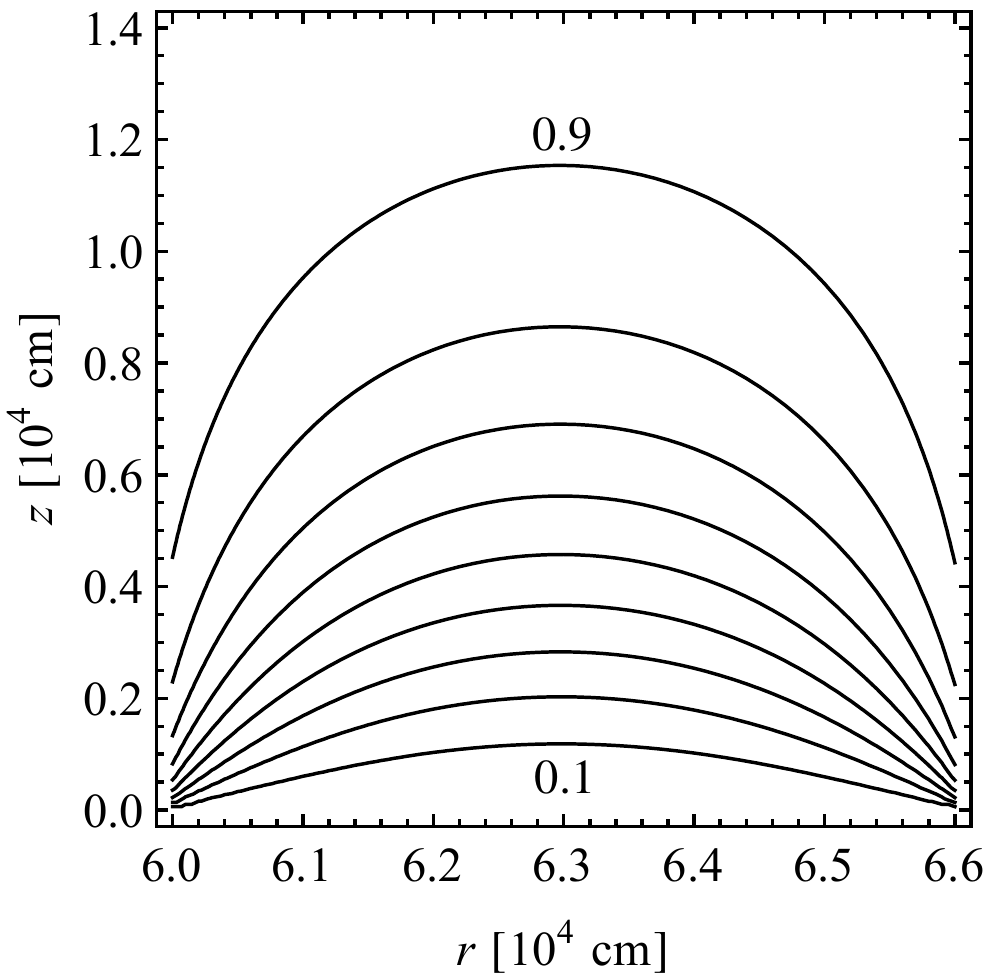}\\a)
%\vspace{5 mm}
\vfill
\includegraphics[height=0.4\textwidth]{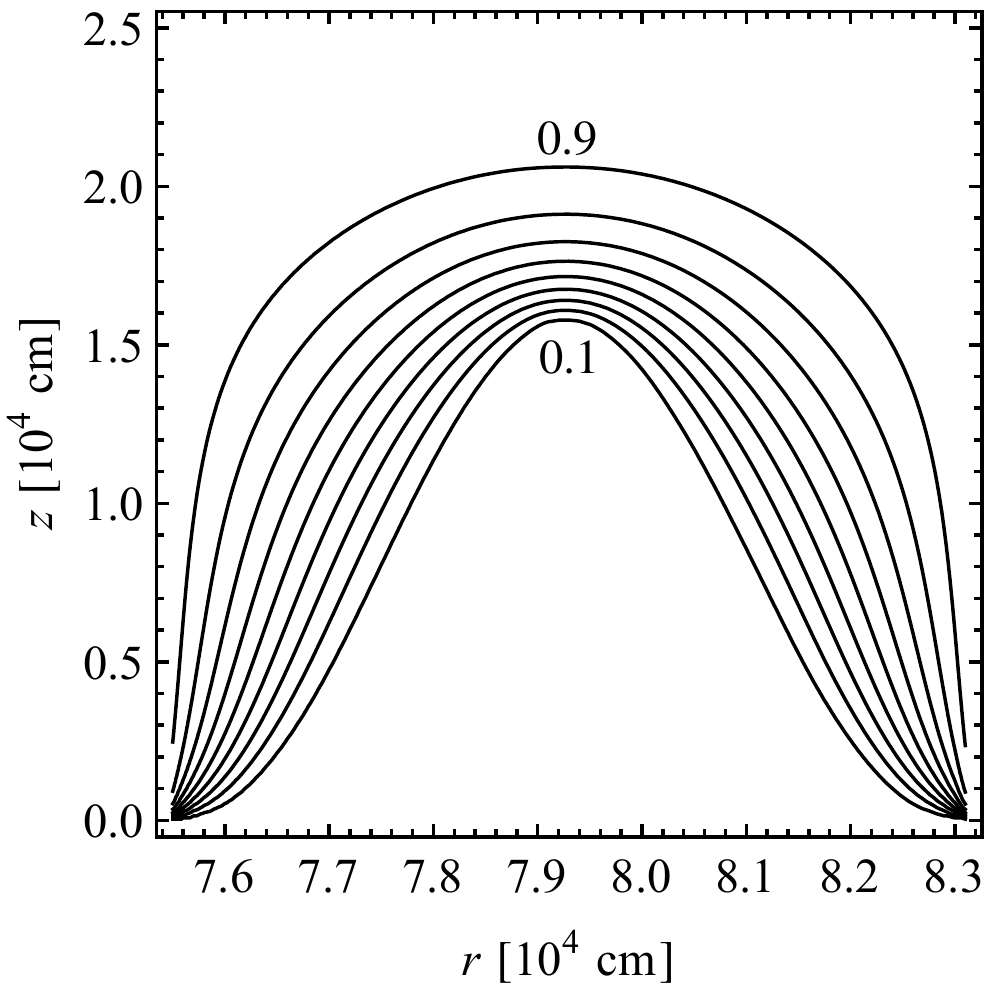}\\b)
\end{multicols}
\hfill
\begin{multicols}{2}
\includegraphics[height=0.4\textwidth]{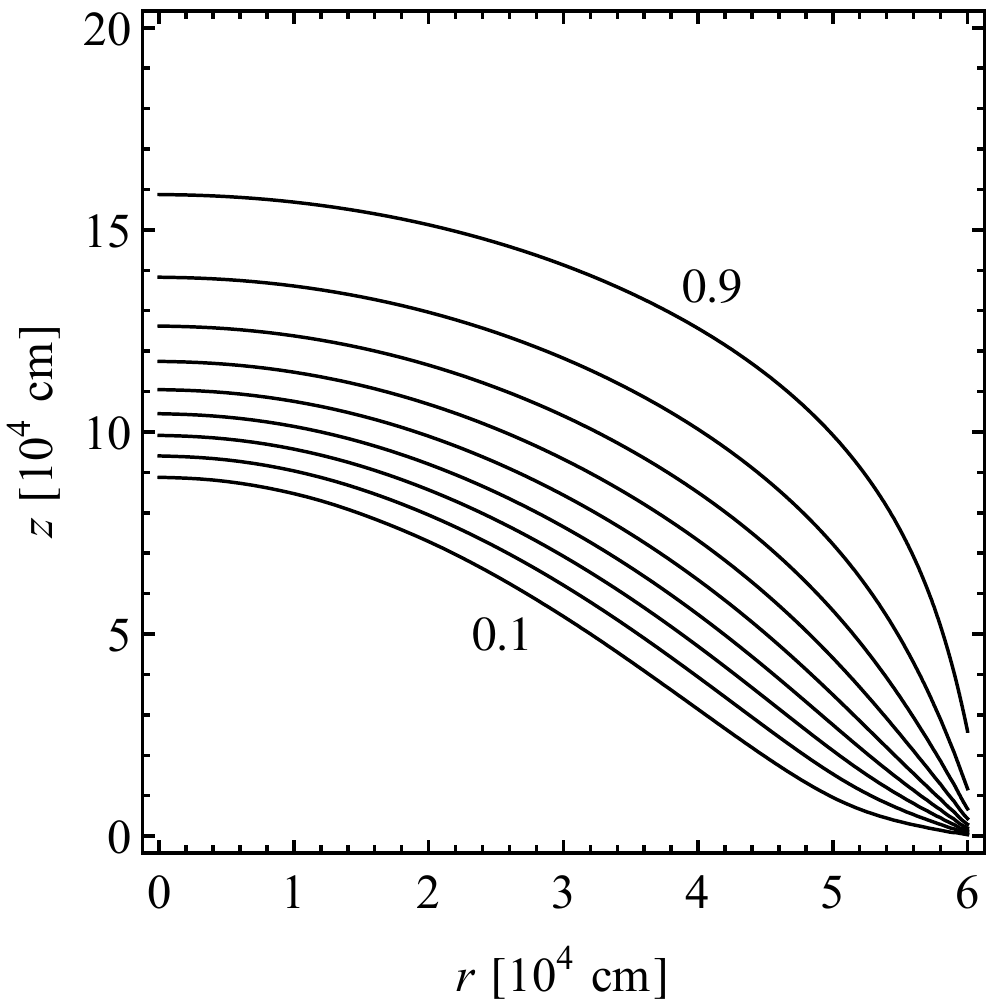}\\c)
%\vspace{5 mm}
\vfill
\includegraphics[height=0.4\textwidth]{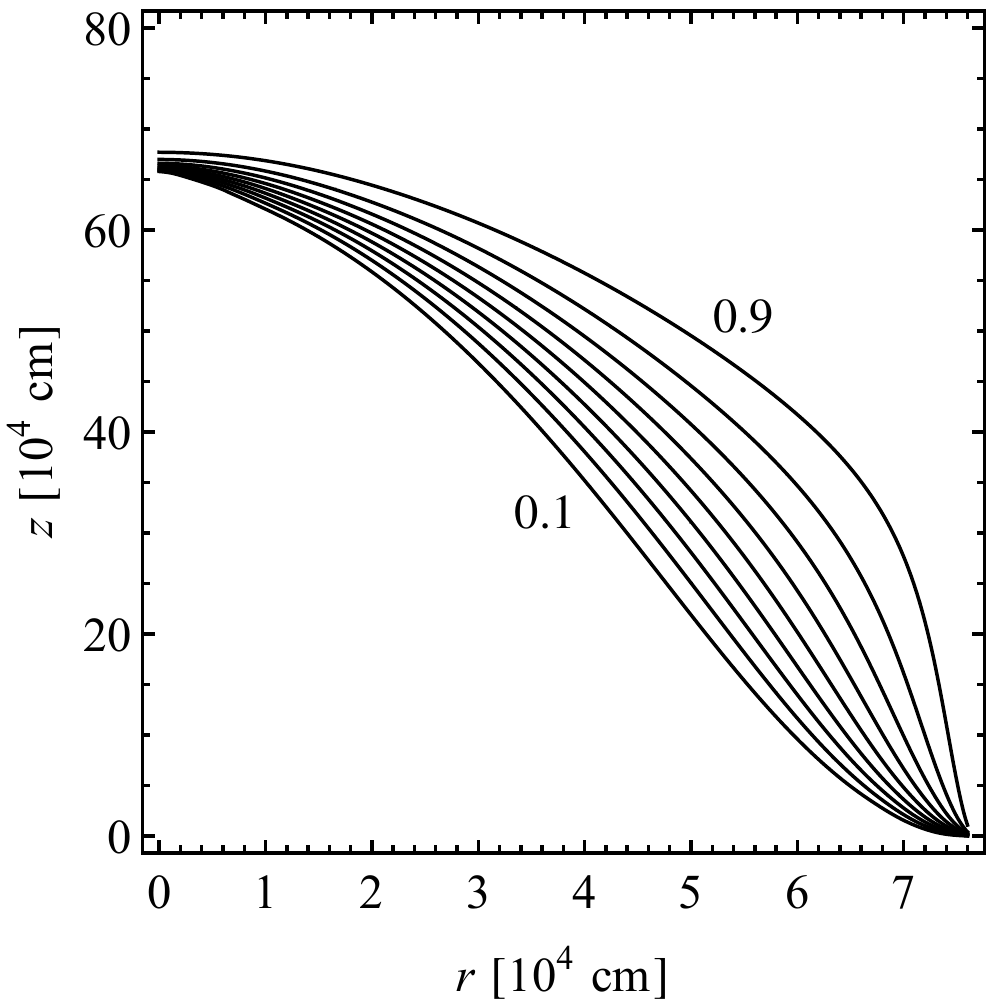}\\d)
\end{multicols}
\label{f:structure}
\caption{Contours of constant $Q$ (0.9 to 0.1, from top to bottom).
Top panels: the hollow cylinder geometry with $b=0.1$ calculated for 
$\dot M_{17}=1$ (a) and $\dot M_{17}=5$ (b).
%a) $\dot{M}=10^{17}~\mbox{\rate}$;
%b) $\dot{M}=5\times 10^{17}~\mbox{\rate}$).
Bottom panels: the filled cylinder geometry 
for $\dot M_{17}=1$ (c) and $\dot M_{17}=5$ (d).
%c) $\dot{M}=10^{17}~\mbox{\rate}$;
%d) $\dot{M}=5\times 10^{17}~\mbox{\rate}$).
}
\label{f:Q}
\end{figure*}

\subsubsection{Boundary conditions}

We are working in cylindrical coordinates $r,\varphi,z$ centred at the columns 
axis and $z=0$ at the neutron star surface.
The initial velocity of the falling matter at high altitude above the surface of the neutron star is $v_0=10^{10}~\mbox{\vel}$. At the cylinder base ($z=0$), the velocity is $v=0$. As a boundary condition at the column side surface we used the
relation between the radial energy flux to the energy density $U$ in the form $F_r(r_0,z)=2cU(r_0,z)/3$, which roughly corresponds to the conditions expected in the scattering
atmospheres in the Eddington approximation.

\subsubsection{Basic equations}

The cylindrical symmetry of the problem makes it essentially two-dimensional. 
The steady-state momentum equation (ignoring gravity, which is very good approximation as discussed e. g. in \cite{BaskoSunyaev76}) for accretion braking by the radiation with energy density $U$ reads
\beq{e:momentum}
  \left(\mathbf{S}\cdot\nabla\right)\mathbf{v}=-\frac{1}{3}\nabla U,
\eeq
where $\bm{S}=\rho\mathbf{v}= \mbox{const}$ is the mass continuity equation. The integration of these equation yields
\beq{e:r-density}
U=3S(v_0-v).
\eeq

Following \cite{1973NPhS..246....1D}, the energy equation can be written as
\beq{e:energy}
\nabla\cdot\mathbf{F}=-\mathbf{S}\cdot\nabla\left(\frac{v^2}{2}\right),
\eeq
where we has neglected the flow of internal energy of the falling matter.

The radiative transfer equation in the diffusion approximation is
\beq{e:transfer}
\mathbf{F} = -\frac{c}{3\varkappa\rho}\nabla U+\frac{4}{3}U\mathbf{v},
\eeq
where the inflow of gas is considered to be stationary.

From equation (\ref{e:transfer}) in cylindrical coordinates we find the components of the radiation energy flux:
\beq{e:Fr}
F_r=\frac{-\frac{c}{3\varkappa_\bot\rho}\frac{\partial U}{\partial r}}
{1+\frac{1}{3\varkappa_\bot\rho}\frac{1}{U}\left|\frac{\partial U}{\partial r}\right|},
\eeq
\beq{e:Fz}
F_z=\frac{-\frac{c}{3\varkappa_\|\rho}\frac{\partial U}{\partial z}-\frac{4}{3}Uv}
{1+\frac{1}{3\varkappa_\|\rho}\frac{1}{U}\left|\frac{\partial U}{\partial z}\right|}.
\eeq
where the coefficients $\left(1+\frac{1}{3\varkappa_\bot\rho}\frac{1}{U}\left|\frac{\partial U}{\partial r}\right|\right)^{-1}$ and $\left(1+\frac{1}{3\varkappa_\|\rho}\frac{1}{U}\left|\frac{\partial U}{\partial z}\right|\right)^{-1}$
are introduced to use the modified diffusion approximation in the optically thin regions where the change of the radiation energy density on the photon mean free path becomes significant (see  \cite{1981A&A....93..255W}).
Equations \eqn{e:momentum} -- \eqn{e:Fz} describe the accretion shock mediated by radiation in the diffusion limit and provide the  structure of the zone where most of the accretion energy responsible for the formation of the emergent spectrum is released. The 'sinking zone' downstream the radiation shock, where the matter settles down subsonically and advects thermal energy towards the NS surface,
should be treated separately (see e.g. \cite{BaskoSunyaev76}). Different physical
effects can appear in this zone, e.g. 'photon bubble oscillations' discussed in \cite{1992ApJ...388..561A,1996ApJ...457L..85K}.

\subsubsection{Equations in dimensionless form}

The convenient  variables are dimensionless velocity of matter squared $Q=v^2/v_0^2$,  the dimensionless column radius $\tilde{r}={\varkappa_T S_0}r/{c}={\tau_T v}r/{c}$ and height $\tilde{z}={\varkappa_T S_0}z/{c}={\tau_T v}z/{c}$, where  $S_0=S(\dot M_{17}=1)=8.79\times 10^6~\mbox{\flux}$ (here and below $\dot M_{17}=\dot M/(10^{17}\hbox{g\,s}^{-1})$ is dimensionless mass accretion rate onto one pole). 
For the assumed value of $r_0(\dot M_{17}=1)$ the dimensionless factors almost exactly equal to $10^{-4}~\mbox{cm}^{-1}$.
For the scattering cross-sections we shall assume $\varkappa_\bot=\varkappa_T$ and $\varkappa_T/\varkappa_\|=10$.

With this notation and dimensionless variables, the system of equations (\ref{e:r-density}), (\ref{e:energy}), (\ref{e:Fr}), (\ref{e:Fz}) is equivalent to one
non-linear partial differential equation  (cf. \cite{1973NPhS..246....1D})
\beq{e:q}
\tilde{K}_\bot\left[\frac{\partial^2 Q}{\partial \tilde{r}^2}+\frac{1}{\tilde{r}}\frac{\partial Q}{\partial \tilde{r}}\right] + \tilde{K}_\|\left[\frac{\varkappa_T}{\varkappa_\|}\frac{\partial^2 Q}{\partial \tilde{z}^2}-8\frac{\partial}{\partial \tilde{z}}\left(\!\sqrt{Q}-Q\right)\right]-\frac{\partial Q}{\partial \tilde{z}}=0,~
\eeq
where the coefficients
$\tilde{K}_\bot=\left(1+\frac{v_0}{6c(1-\sqrt{Q})}\left|\frac{\partial Q}{\partial \tilde{r}}\right|\right)^{-1}$ and   $\tilde{K}_\|=\left(1+\frac{\varkappa_T}{\varkappa_\|}\frac{v_0}{6c(1-\sqrt{Q})}\left|\frac{\partial Q}{\partial \tilde{z}}\right|\right)^{-1}$
are calculated at each consecutive iteration using values of $Q$ obtained in the previous iteration.

%\subsubsection{Numerical procedure}

Equation (\ref{e:q}) was solved numerically using the finite difference approximation method. A time-relaxation scheme was used. In this approach, the equation in the form $L[Q]=0$, where $L$ is the corresponding differential operator, has been replaced by a parabolic equation of the form $\frac{\partial Q}{\partial t}-L[Q]=0$. The steady-state solution of the last equation coincides with the solution of the original elliptic equation. An explicit finite-difference scheme was used to solve the parabolic equation.

\subsection{Results of numerical calculations}
\label{s:res}

Numerical simulations were carried out for accretion rates onto one pole in the range
$10^{17}-1.2\times10^{18}~\mbox{\rate}$
%and $10^{17}\div 7\times 10^{17}~\mbox{\rate}$
for the hollow cylinder geometry and the filled cylinder geometry. The column structure at different accretion rates is shown in \Fig{f:Q}. Note that
at a given mass accretion rates and the NS magnetic field, the height of the braking zone in the hollow cylinder case is smaller than that in the case of the filled cylinder, as expected from the dependence of the radiation diffusion time on the
wall thickness $\propto b$. Clearly, in very geometrically thin walls, $b/r_0\ll 1$,
the flow will be
optically thin as well (at least for the extraordinary mode with angle-independent
scattering cross-section in the strong magnetic field). This may suggest that
even if the flow was initially confined within a thin layer, near the NS surface
the huge energy release may change the character of the flow. This possibility is worth investigating further.

Guided by the velocity braking in a radiation-dominated shock, $v\approx1/7 v_0$, 
we define the 'height' of the accretion mound $z_0(\dot{M})$ as the distance 
at the centre of the structure where the gas flow velocity is reduced to one seventh of the initial value. The (dimensionless) height calculated for different accretion rates were fitted by a power law
\beq{}
\tilde{z}_0=\beta\dot{M}^\alpha_{17},
\eeq
where $\tilde{z}_0=\varkappa_T S_0 z_0/c$. We found that in both geometries
$\alpha\sim 1$ and $\beta>0$ ($\alpha\simeq 1.1$, $\beta\simeq 0.3$ for the case of hollow cylinder 
and  $\alpha\simeq 0.9$, $\beta\simeq 13.5$ for the filled cylinder).
Therefore, the numerical solution behaves in agreement with the qualitative 
physical considerations about the structure of optically thick accretion mounds described above.

\section{Change in the hardness ratio from accretion columns}
\label{s:spectrum}

\subsection{Saturated Compton spectrum of sidewall emission}

\begin{figure}
\includegraphics[width=0.48\textwidth]{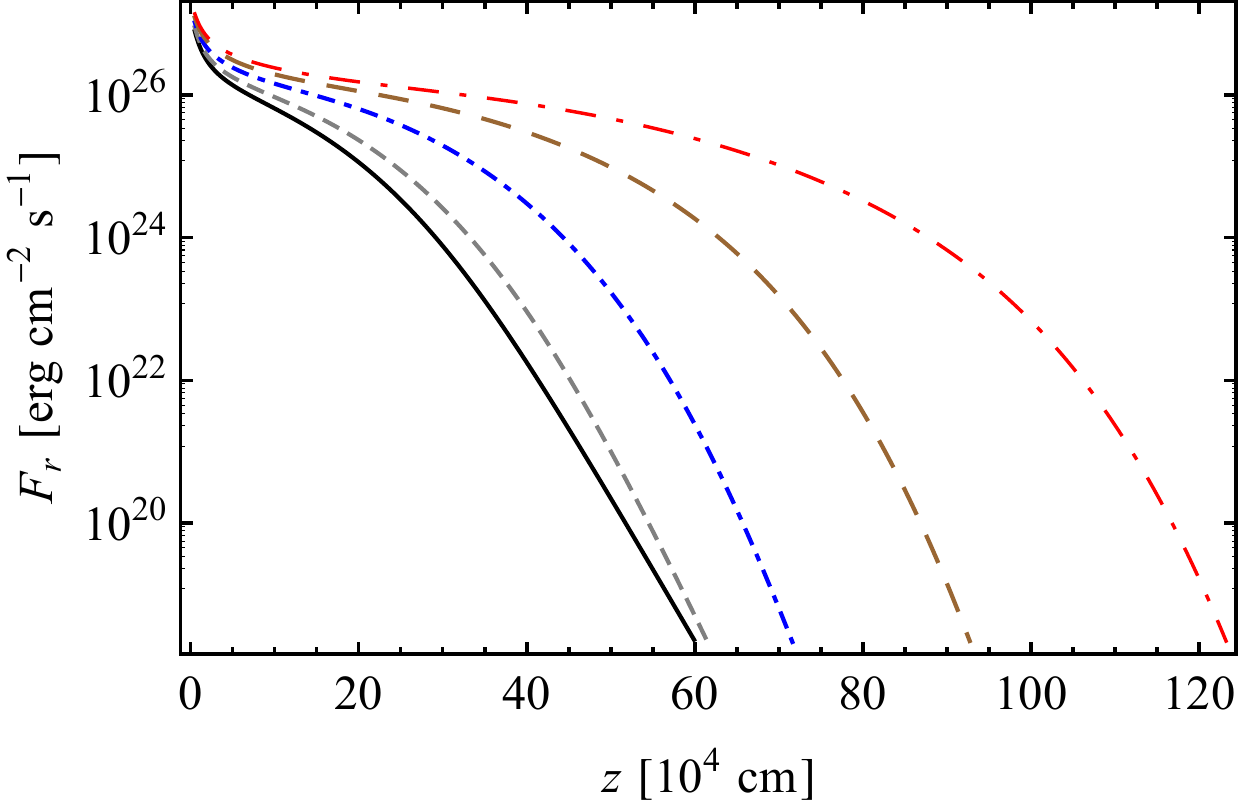}
\caption{
The radial energy flux $F_r(\tau=2/3,z)$ calculated along the height
of a filled cylinder column 
at different mass accretion rates. Half of this flux (extraordinary photons
only) escapes 
sidewall. The lines from bottom to up correspond to 
$\dot M_{17}=2$, 3, 5, 8 and 12, respectively.
}
\label{f:F_r}
\end{figure}

\begin{figure*}
\includegraphics[height=0.34\textwidth]{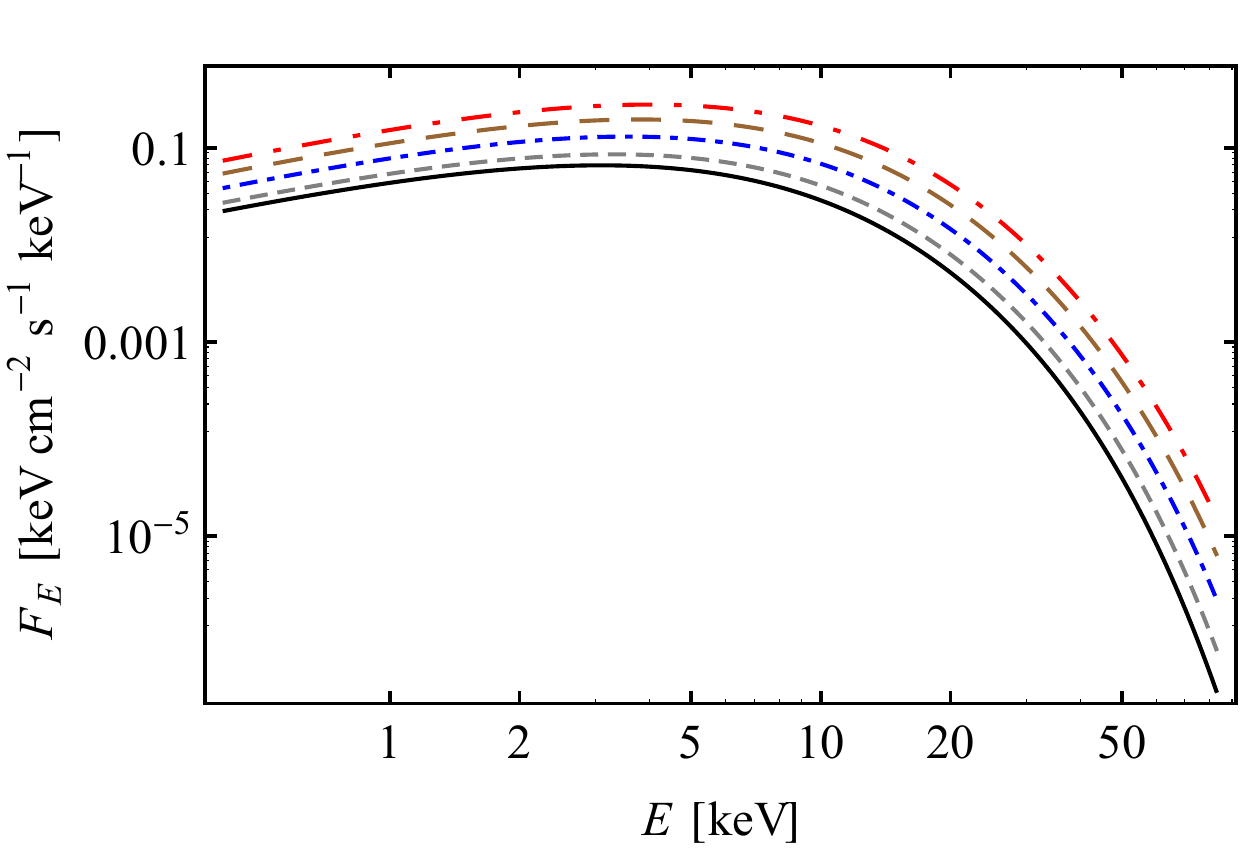}
\hfill
\includegraphics[height=0.34\textwidth]{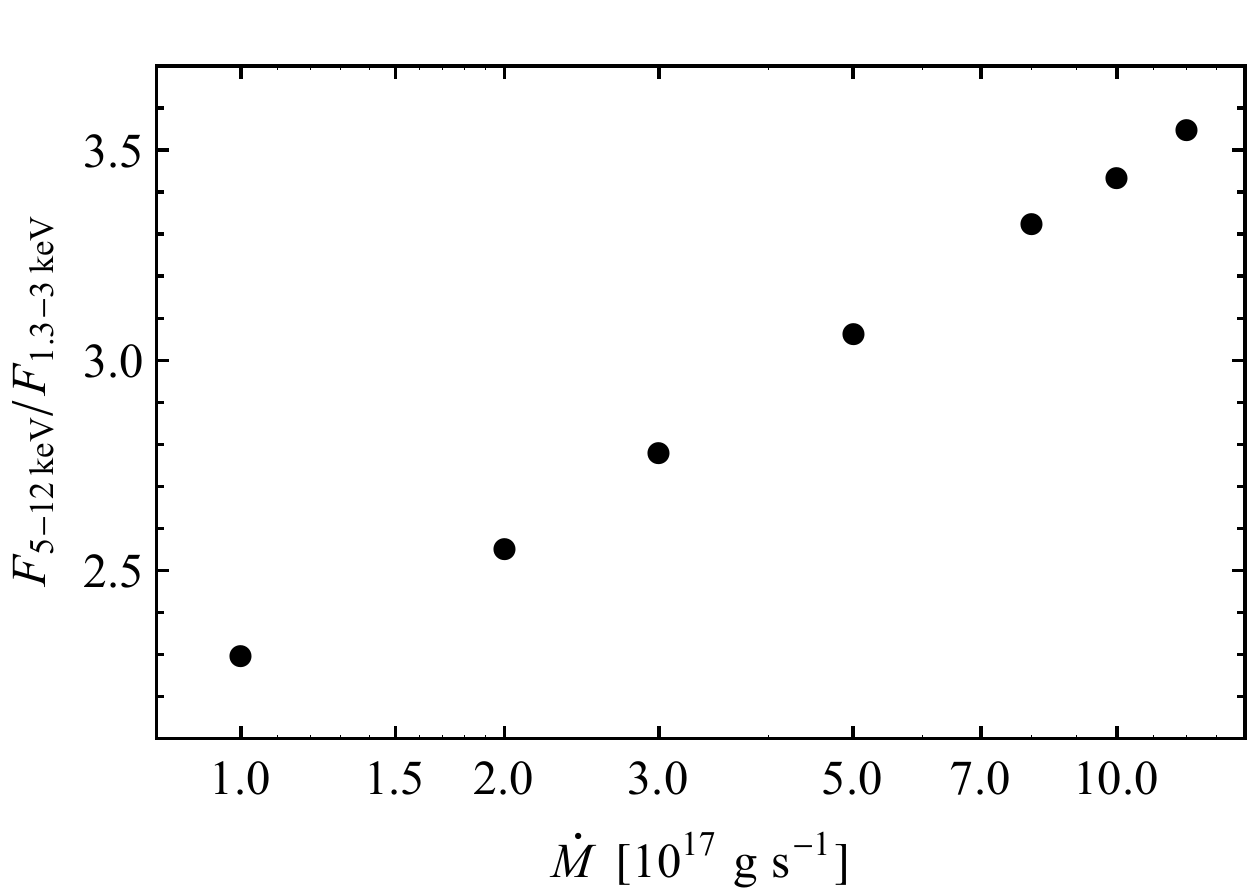}
%article_xlog_2.pdf}}
\caption{
Left:
The spectrum of sidewall emission from optically thick filled cylinder accretion column
\eqn{e:IF}
for mass accretion rates $\dot M_{17}=2$, 3, 5, 8 and 12 
(from bottom to up, respectively).
Right: The spectral hardness ratio $HR$
as a function of mass accretion rate.  
}
\label{f:col_spec}
\end{figure*}

The LTE treatment of the spectrum of the sidewall emission from
accretion columns is clearly a strong oversimplification. To take into
account the scattering on electrons in the optically thin boundary of the column, we
can use the model calculations of the radiation transfer problem
in a semi-infinite plane-parallel atmosphere with strong magnetic field.
In this case the emerging
spectrum will be formed in the saturated Compton regime, with the mode 1 photons (extraordinary, i.e.
polarized perpendicular to the $\bm{k}-\bm{B}$-plane,
where $\bm{k}$ is the photon wavevector and $\bm{B}$ is the magnetic field vector)
predominantly escaping through sidewalls \citep{1986Ap.....25..577L}. The intensity
of mode 2 (ordinary, i.e. polarized in the $\bm{k}-\bm{B}$ plane) photons is comparable to
that of mode 1 photons only at large angles to the normal, $\mu'\sim 0$,
and thus insignificantly contribute to the total flux from the unit surface
area due to the
geometrical factor.

In this regime,
the specific intensity of extraordinary photons 
at energies far below the cyclotron resonance can be written to an accuracy of
a few per cents as \citep{1986Ap.....25..577L}
\beq{e:I1}
I_\nu'=\frac{2\sqrt{3}}{5}(1+2\mu')\frac{h\nu \nu_g^2}{c^2 \tau_0}e^{-\frac{h\nu}{kT}}\,,
\eeq
where $\mu'=\cos\theta'$ is the angle between the normal to the atmosphere and
the escaping radiation direction in the plasma reference frame, $\tau_0$ is the characteristic optical depth of the problem, $\nu_g=\frac{eB}{2\piup m_ec}$ is the electron gyrofrequency in the magnetic field. By expressing the characteristic optical depth $\tau_0$ through
the emergent radiation flux $\Phi$ using Eqs. (34-37) from \cite{1986Ap.....25..577L},
we find
\beq{e:IF}
I_\nu'=\frac{3}{10\piup}\myfrac{h\nu}{kT}^2\frac{\Phi}{\nu}e^{-\frac{h\nu}{kT}}\,.
\eeq

Therefore, we are in the position to calculate the emerging spectrum from the column
in this approximation using the solutions for the column structure obtained above.
To do this, it is sufficient to substitute the total radial energy flux
at each height of the column $\Phi\to F_r(z)/2$, and temperature $T\to T(z)$ estimated
deep inside the column (at the optical depth $\sim 1$),
so that $T(z)=(U(z)/a_r)^{1/4}$). Note here that at large optical depth 
the number density of ordinary and extraordinary photons are equal, with the
extraordinary photons becoming overwhelmingly dominant in the optically thin outer
layers. Therefore, only half of the radial energy flux, $F_r(z)/2$, should
be taken into account in the spectral calculations. The second half of
the radial flux (in ordinary photons) is admixed to the vertical energy flux
component, $F_z$. Some fraction of this energy flux can propagate (mostly 
as ordinary photons) along the magnetic field into the optically thin upper
part of the column;  we do not calculate their fate in this paper and
restrict ourselves by this quantitative note.
The radial energy flux $F_r(z)$ from 
the filled column is shown in Fig. \ref{f:F_r} for different mass accretion
rates.   

The resulting
spectra of the sidewall emission from 
the optically thick accretion column integrated over the column height $z$
is shown in Fig. \ref{f:col_spec} (left panel) for the case of the filled column and different
$\dot M$. The spectral hardness ratio 
\beq{}
HR=\frac{\int\limits_{5~\mathrm{keV}}^{12~\mathrm{keV}}F_\nu d\nu}
{\int\limits_{1.3~\mathrm{keV}}^{3~\mathrm{keV}}F_\nu d\nu},
\eeq
is shown in the right panel of this figure as a function of $\dot M$. It is
seen that the hardness ratio of the spectrum of the column monotonically 
increases with mass accretion rate. 

\subsection{Account for the reflected component}

We have seen that the saturated-Comptonized spectrum of sidewall
emission from optically thick accretion column gets harder with increasing $\dot M$,
contrary to what is observed and discussed above. However, an important feature of
accretion in X-ray pulsars should be added at this point. 
Namely, we should take into account the fact that the electrons in the optically thin part of the column are moving
with high velocity $v_0\sim 1/3 c$, 
and thus the emission beam of the column
\eqn{e:I1} should be Doppler-boosted towards the neutron star surface and reflected from the neutron star atmosphere, as discussed
in \cite{1988SvAL...14..390L} and \cite{2013ApJ...777..115P}. The reflection coefficient in the case of single Compton electron scattering
in strong magnetic field is determined by the ratio $\lambda(\nu)=\kappa_{sc}/(\kappa_{sc}+\kappa_{abs})$, where $\kappa_{sc}$ and $\kappa_{abs}$
is the scattering and absorption coefficients for photons in the strong magnetic field, respectively.
The reflection coefficient (X-ray albedo) is calculated in the Appendix
\ref{s:appendix} and is presented in Fig.
\ref{f:albedo}. Note that the calculated X-ray albedo turns out to be very similar for
both extraordinary and ordinary photons and is virtually insensitive to
the photon incident angle to the magnetic field. 
Everywhere below we shall use X-ray albedo for extraordinary photons only. 
It is seen that soft photons are mostly
absorbed, while hard photons are reflected by the neutron star atmosphere.
Therefore,  the reflected spectrum  is harder than the incident one.

\begin{figure}
\includegraphics[width=0.48\textwidth]{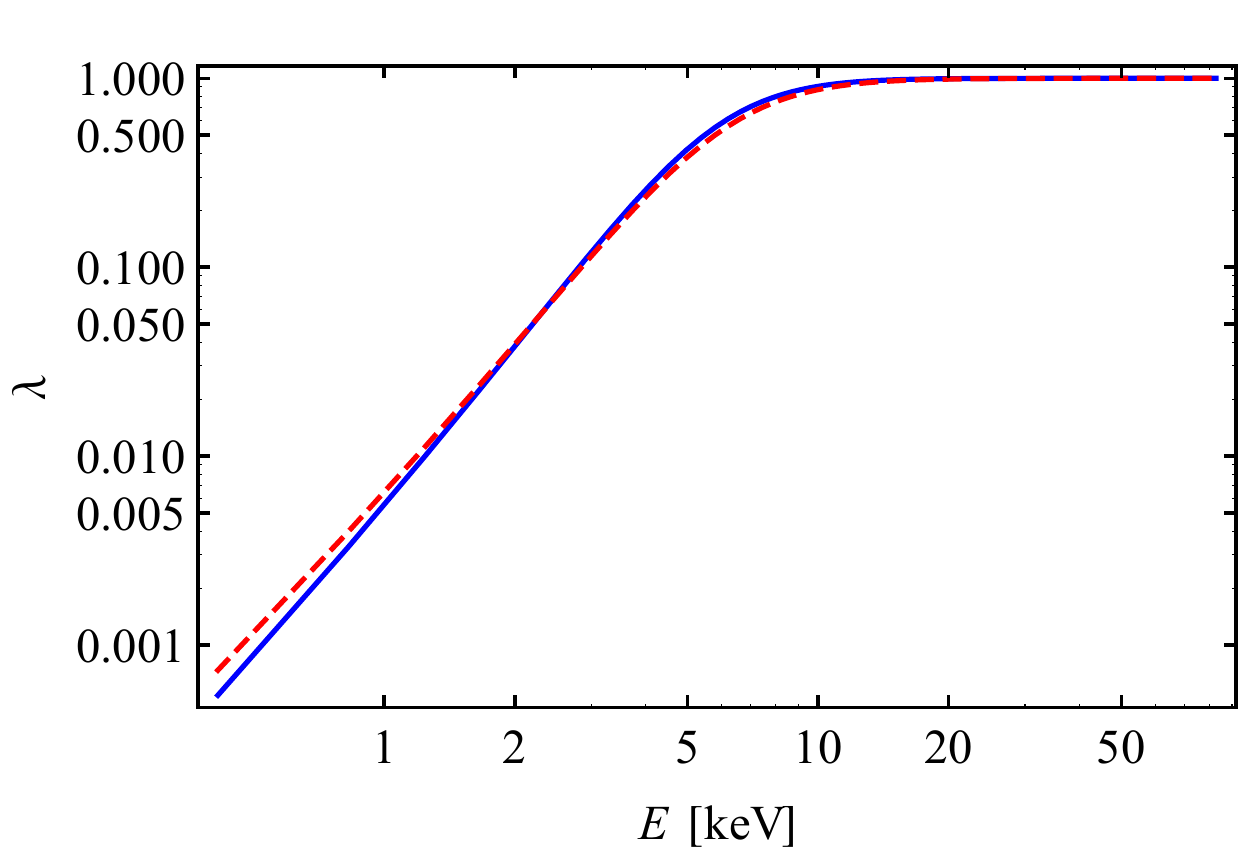}
\caption{
The single-scattering Compton X-ray albedo from a plane-parallel
neutron star atmosphere with strong magnetic field. The electron number
density is $n_e=5\times 10^{25}$~cm$^{-3}$, electron temperature $T=3$~keV, 
magnetic field $B=3\times 10^{12}$~G, 
photon incident angle to the magnetic field $\piup/4$.  
Extraordinary and ordinary 
photon spectra are shown by the solid and dashed lines, respectively.  
The cyclotron line is ignored.
}
\label{f:albedo}
\end{figure}

Let $\alpha$ be the angle between the photon wavevector and 
the plasma bulk velocity vector.
Due to the boosting, part of radiation from the column
will be intercepted by the neutron star surface and reflected.
The critical angle $\alpha^*$ within which the radiation from the
height $z$ above the surface will be intercepted
by the neutron star (in the Schwarzschild metric) is \citep{2013ApJ...777..115P}
\beq{}
\sin \alpha^*=\frac{R_{NS}}{R_{NS}+z}\sqrt{\frac{1-r_S/(R_{NS}+z)}{1-r_S/R_{NS}}},
\eeq
where $r_S=2GM/c^2$ is the Schwarzschild radius. The difference between the polar axis and the column cylinder side can be neglected for $r_0\ll R_{NS}$.

Let $\Sigma$ be the surface of the column, $\Sigma'(\varphi, z)$ be a point on the surface and
$d\Sigma=r_0 d\varphi dz$ be the elementary area.
The flux from the surface element at frequency $\nu$ (in conventional notations) reads
\beq{ell}
f_\nu(\Sigma')=\frac{dE(\Sigma')}{dtd\nu r_0d\varphi dz}
\eeq
and can be separated into three parts in accordance to the angle relative to the observer.
Photons escaping with angles $\alpha>\alpha^*$ to the column are directly seen
as the proper column radiation $f_\nu^{col}$.
The second part includes the radiation $f_\nu^{ref}$ intercepted by the
neutron star and reflected
from the neutron star surface with X-ray albedo $\lambda (\nu)$.
The third part $f_\nu^{abs}$ is the radiation absorbed and re-radiated 
by the neutron star atmosphere. Then we can write:
\begin{eqnarray}
\label{dL}
&f_\nu(\Sigma')= \int\limits_{-1}^1\frac{d f_\nu(\Sigma')}{d \cos \alpha} d\cos\alpha \nonumber\\
&=\int\limits_{-1}^{\cos\alpha^*}\frac{d f_\nu(\Sigma')}{d \cos \alpha} d\cos\alpha + \int\limits_{\cos\alpha^*}^1\frac{d f_\nu(\Sigma')}{d \cos \alpha} d\cos\alpha \nonumber\\
&=\int\limits_{-1}^{\cos\alpha^*}\frac{d f_\nu(\Sigma')}{d \cos \alpha} d\cos\alpha\nonumber\\
&+\lambda(\nu)\int\limits_{\cos\alpha^*}^1\frac{d f_\nu(\Sigma')}{d \cos \alpha} d\cos\alpha +
(1-\lambda(\nu))\int\limits_{\cos\alpha^*}^1\frac{d f_\nu(\Sigma')}{d \cos \alpha} d\cos\alpha\nonumber\\
&=f_\nu^{col}(\Sigma')+f_\nu^{ref}(\Sigma')+f_\nu^{abs}(\Sigma').
\end{eqnarray}
The integrands \citep{2013ApJ...777..115P} in the cylindrical coordinates
with account for the axial symmetry ($I_\nu(\Sigma')=I_\nu(z)$, giving by (\ref{e:I1}))
have the form:
\beq{dLdcos1}
\frac{df_\nu(\Sigma')}{d\cos\alpha}
=I_\nu(z)\frac{2D^3}{\gamma}\sin\alpha\left(1+\frac{\piup D}{2}\sin\alpha\right),
\eeq
where the Doppler factor is $D=1/(\gamma(1-\beta\cos\alpha))$,
$\gamma=1/\sqrt{1-\beta^2}$ is the plasma Lorentz factor and $\beta=v/c$.
Here we have taken into account that the specific flux (per unit frequency interval) is $D$ times as small as the integral flux. 
Then for the total radiation from the column at frequency $\nu$ we obtain:
\begin{eqnarray}\label{L}
&L_\nu = \iint\limits_\Sigma f_\nu(\Sigma')d\Sigma
=2\piup r_0\int\limits_0^{z_{max}}f_\nu(z)dz\nonumber\\
&=2\piup r_0\int\limits_0^{z_{max}}(f_\nu^{col}(z)+f_\nu^{ref}(z)+f_\nu^{abs}(z))dz\nonumber
\\
&=L_\nu^{col}+L_\nu^{ref}+L_\nu^{abs},
\end{eqnarray}
where $z_{max}$ is the upper limit of the computational area.
The observed X-ray flux from one column is 
$F_\nu=(L_\nu^{col}+L_\nu^{ref})/(4\piup d^2)$, where $d$ is
the distance to the source. For the illustrative purposes, we assume
$d=5$~kpc. 
Taking into account of the radiation absorbed by the neutron star 
atmosphere $L_\nu^{abs}$ will add more soft photons to the total spectrum.

\begin{figure*}
\includegraphics[height=0.34\textwidth]{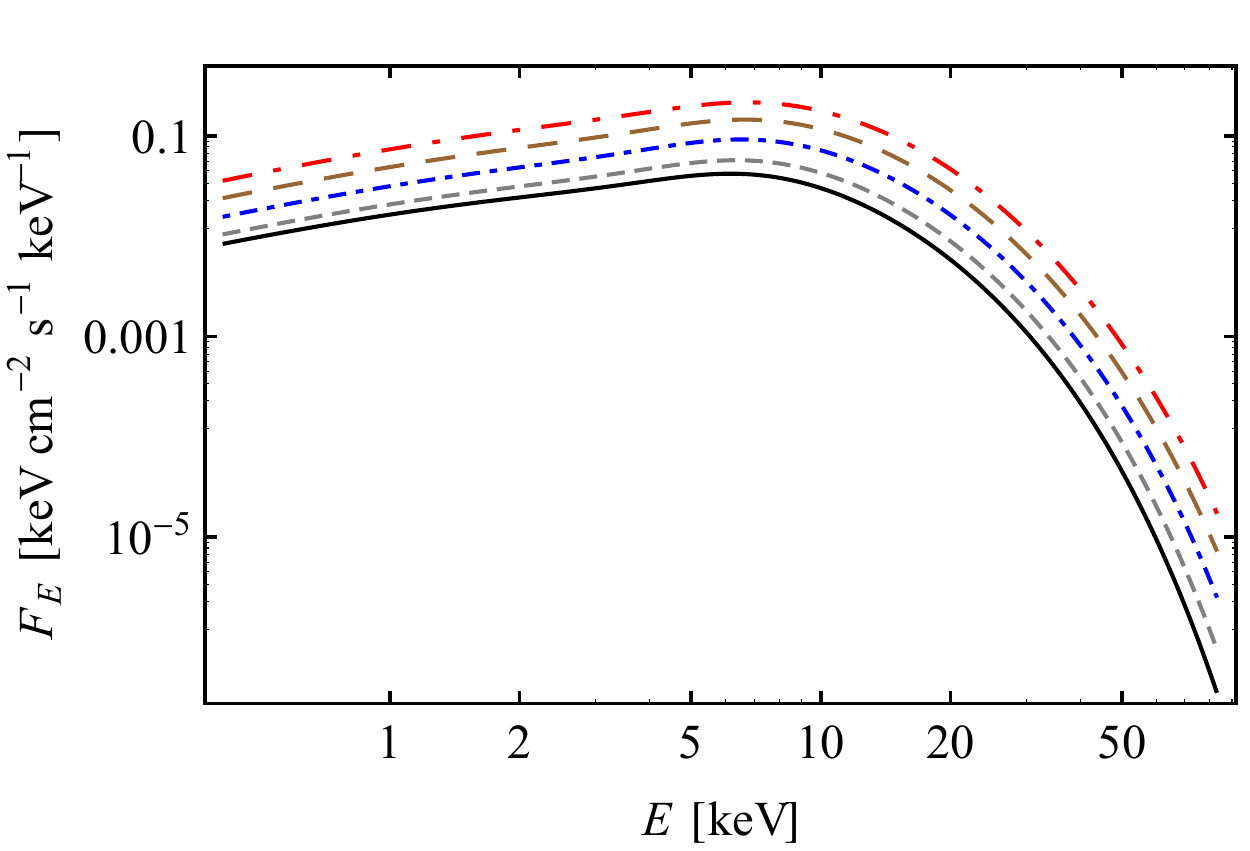}
\hfill
\includegraphics[height=0.34\textwidth]{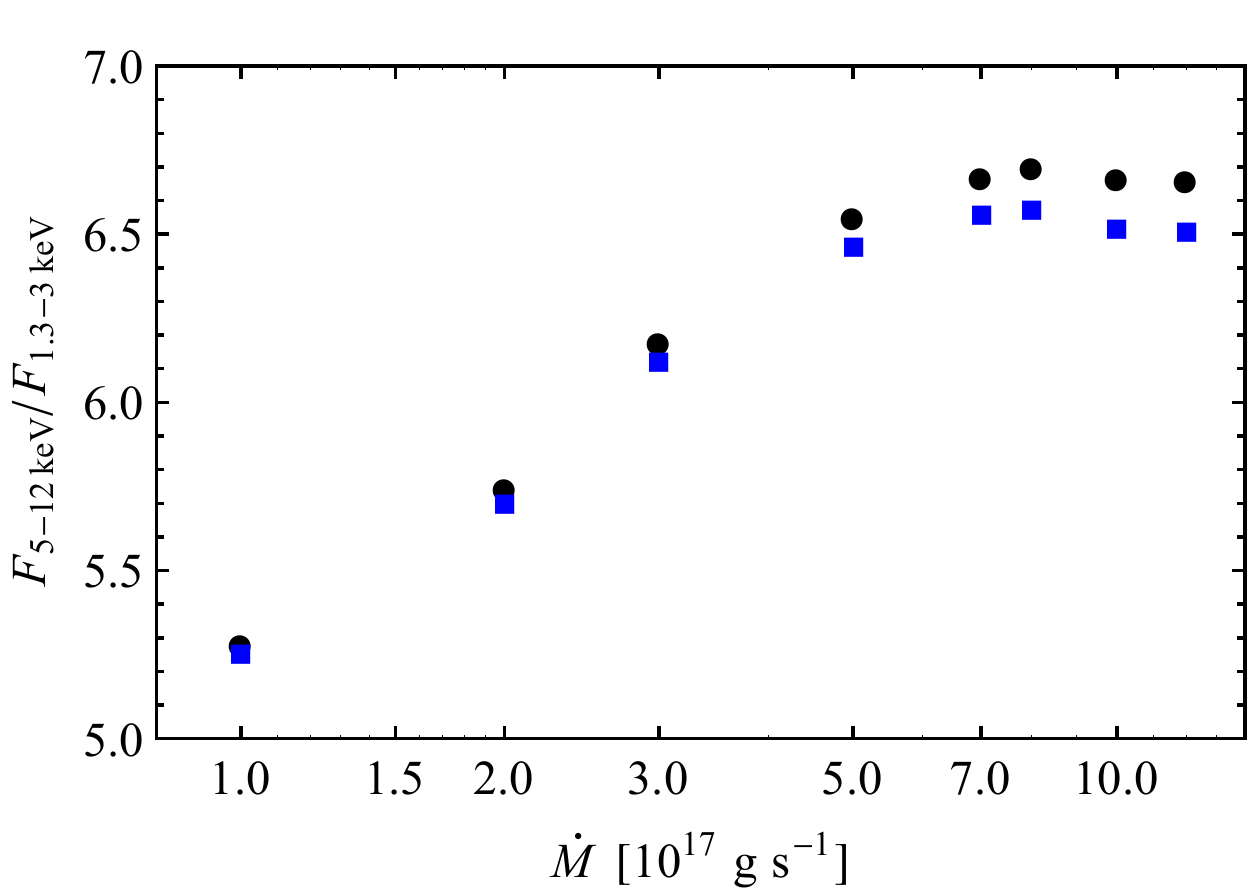}
\caption{
Left: the total spectrum of direct sidewall and reflected 
from the neutron star atmosphere from the optically thick filled
accretion column for mass accretion rates $\dot M_{17}=2$, 3, 5, 8 and 12 
(from bottom to up, respectively).
Right: the hardness ratio $HR$ of the total spectrum. 
Shown are calculations for the neutron star radius $R_{NS}=10$~km (squares) and $R_{NS}=13$~km
(circles).  
}
\label{f:total}
\end{figure*}

Since we are observing both the direct and reflected radiation from the column, 
the total change in the hardness of the spectrum will depend on the fraction of the
reflected radiation in the total flux. This, in turn, depends on the
height of the column. We have calculated the total flux as the sum of the direct and reflected component as a function of $\dot M$ with taking into account the  
photon ray propagation in the Schwarzschild metric of the neutron star
with a fiducial mass $M=1.5 M_\odot$ and radius $R=10$ and 13~km.
The magnetic field of the neutron star is set to $3\times 10^{12}$~G, so that the
CRSF energy is about 35 keV.
The result is presented in Fig. \ref{f:total}. It is seen that the hardness ratio
$HR$ first increases with the mass accretion rate (X-ray luminosity), but starting
from $\dot M_{cr}\sim (6-8)\times 10^{17}$~g~s$^{-1}$ it gets saturated and even decreases.
This is in agreement with observations shown in Fig. \ref{f:hardness}.

\begin{figure}
\includegraphics[width=0.48\textwidth]{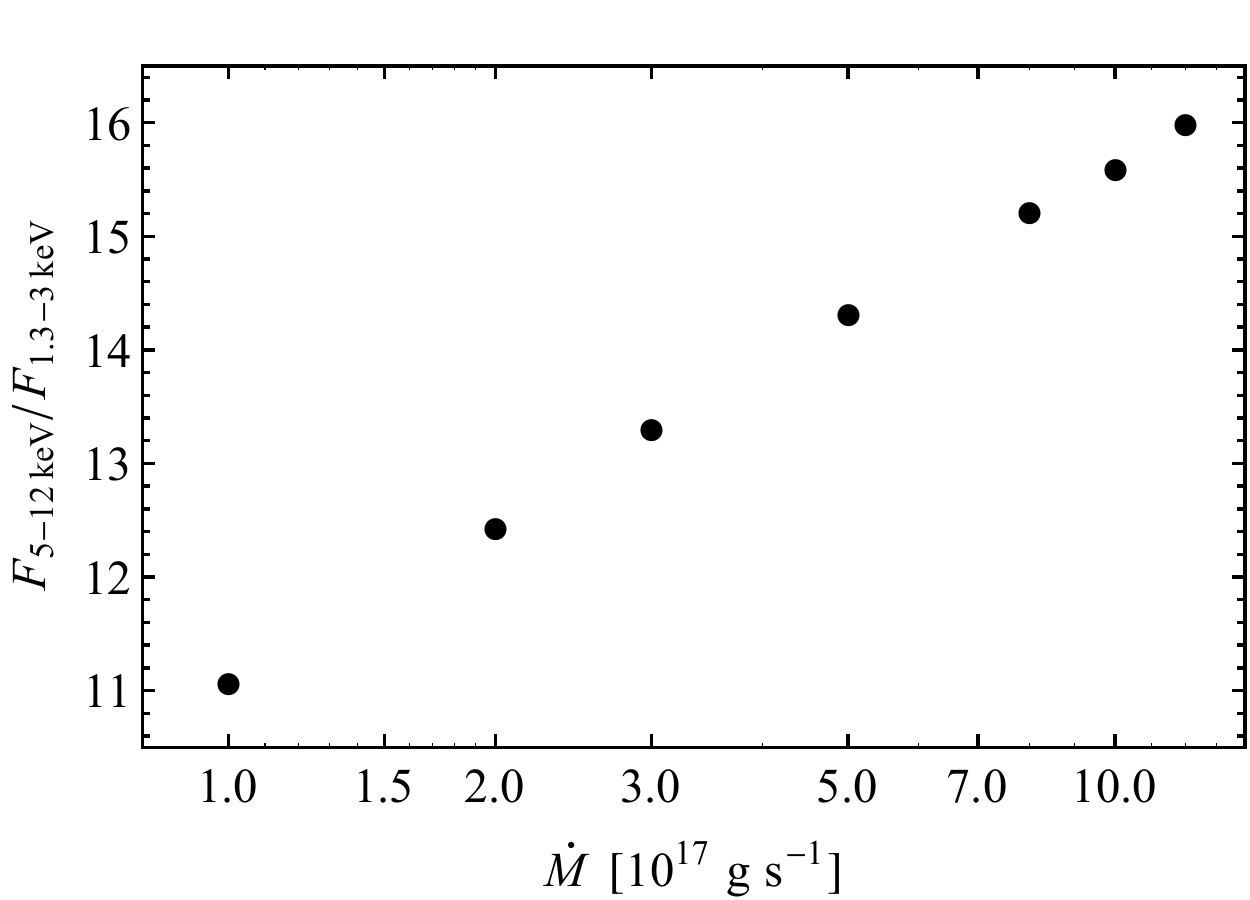}
\caption{
Hardness ratio of the total emission (direct plus reflected from the 
neutron star atmosphere) from a hollow cylinder accretion column 
with $b=0.1r_0$ as a function of mass accretion rate $\dot M$. 
}
\label{f:hr_hol}
\end{figure}

For the hollow cylinder accretion columns which have smaller height
(see Fig. \ref{f:Q}), the fraction of the reflected component 
in the total emission does not virtually change with mass accretion rate within the calculated range, therefore the hardness ratio of the total spectrum monotonically
increases with $\dot M$ in this range (see Fig. \ref{f:hr_hol}). Note also 
that in this case the spectrum is harder than in the case of the filled column because
the reflected (harder) emission dominates in the total spectrum.

\section{Summary and conclusions}

In this paper, using RXTE/ASM archival data we studied the
behaviour of the spectral hardness ratio as a function of X-ray luminosity at different accretion
rates in a sample of six transient X-ray
pulsars (EXO 2030+375, GX 304-1, 4U 0115+63, V 0332+63, A 0535+26 and
MXB 0656-072). In all cases, the hardness ratio
is found to increase with X-ray flux at low luminosities and then to
saturate or even slightly to decrease above some critical
luminosity of a few times $10^{37}$~erg~s$^{-1}$
This behaviour confirms earlier findings by
\citet{Reig:Nespoli:13}, and \citet{Klochkov:etal:11}
and correlates with the behaviour of the CRSF energy with flux
in the CRSF sources.

%We propose a simple physical explanation to the observed correlation in terms of 
%different regimes of accretion onto magnetized neutron stars. 

At low mass accretion rates and low X-ray luminosities, 
the braking of infalling matter is expected to be mediated by Coulomb interactions 
\citep{1969SvA....13..175Z, 1993ApJ...418..874N, Becker_ea12} and occurs close to the neutron star surface, 
essentially at a height of the homogeneous atmosphere above the neutron star surface, 
without the formation of an accretion column. When the X-ray luminosity increases above some 
critical value $\sim 10^{37}$~erg s$^{-1}$ 
\citep{1973NPhS..246....1D, BaskoSunyaev76, 1981A&A....93..255W, 2015MNRAS.447.1847M}, 
the radiation pressure of generated photons decelerates the infalling matter. 
An optically thick accretion column arises with the height increasing almost linearly with accretion rate.

Therefore, at low luminosities $< L^*\sim 10^{37}$~erg s$^{-1}$ 
the emerging X-ray spectrum is essentially 
formed by ordinary photons in a magnetized optically thin atmosphere. The hardness of the spectrum 
is determined by the Comptonization $y$-parameter, $y=(kT/m_ec^2)\tau$, 
which increases with accretion rate. This implies that in this case the hardness of the emerging radiation will increase with luminosity.

At higher luminosities an optically thick accretion column above the
magnetic pole arises, and thermal radiation generated deep inside
the column escapes sidewall predominantly as extraordinary
photons. Since at energies far less than the CRSF energy 
the electron scattering cross-section strongly decreases with
frequency, the spectrum of the emergent continuum radiation will be
softer than the diluted Wien spectrum expected 
to be formed in the saturated Compton
regime, $\sim \nu e^{-h\nu/kT}$ \cite{1986Ap.....25..577L}.  

We have calculated the structure of axially symmetric optically thick accretion columns in the diffusion approximation for radiation transfer
for different mass accretion rates. Two accretion column geometries, filled cylinder and hollow cylinder, were considered. 
In both cases we have found the expected almost linear growth of the column height with mass accretion rate.
The continuum X-ray spectrum of the sidewall emission in both geometries gets
harder with increasing mass accretion rate. Taking into account of
the radiation reflected from the neutron star atmosphere allowed us to
reproduce the observed saturation of the continuum hardness. This is purely
geometrical effect: with increasing mass accretion rate the height of the
column increases, so does the fraction of the reflected radiation. However,
starting from some height, the fraction of the reflected radiation in the
total flux stops
increasing and starts decreasing. Since the hardness of the reflected
radiation is higher than that of the incident emission  (because of the
strong energy dependence of the scattering and absorption coefficients in
the strong magnetic fields), the hardness of the total spectrum (direct 
sidewall emission from the accretion column plus reflected radiation from
the neutron star atmosphere) saturates (and even can slightly decrease) beyond 
some characteristic mass accretion rate $\sim (6-8)\times 10^{17}$~erg s$^{-1}$ for the 
typical
neutron star parameters: $M_{NS}=1.5 M_\odot$, $R_{NS}=10$~km and 13~km, $B=3\times
10^{12}$~G. 

In the case of the hollow cylinder accretion column geometry with a fiducial
wall thickness of $0.1$ the outer column radius $r_0$, 
we have not found the spectral hardness ratio saturation with increasing
mass accretion rate up to the maximum value of our calculations 
$1.2\times 10^{18}$~g s$^{-1}$. Apparently, the height of the column in this case is 
too low, the hard reflected component dominates at all mass accretion rates, 
and the (softer) direct sidewall emission remains subdominant in the total
spectrum. 
%even 
%at mass accretion rates as high $1.2\times 10^{18}$~g~s$^{-1}$. 

Therefore, our model calculations lead to the following conclusions. 

1. The spectral hardening in X-ray pulsars with positive CRSF energy
dependence on X-ray flux can be explained by increasing of the
Comptonization parameter $y$ in the slab atmosphere of the accretion mound.
It is in this regime that 
positive correlation of the cyclotron line is observed, for example, 
in Her X-1 \citep{2007A&A...465L..25S}.

2. At high accretion rates, the radiation-supported optically thick 
accretion column grows above the polar cap,
the sidewall emission from the column is formed by extraordinary photons
in the saturated Compton  regime. The spectrum of this emission gets
harder with increasing mass accretion rate. 

3. With further increasing mass accretion rate the height of the column
increases, such that the fraction of radiation reflected from the neutron
star atmosphere starts decreasing. As the
reflected radiation is harder than the incident one, the
spectrum of the total emission 
(direct plus reflected) stops hardening (and even becomes slightly softer). 
%For the fiducial neutron star parameters
%($M_{NS}=1.5 M_\odot$, $R_{NS}=10$~km and 13~km, $B=3\times 10^{12}$~G) 
This happens at the mass accretion rate onto one pole $\dot M_{cr}\sim (6-8)\times 10^{17}$~g s$^{-1}$, in qualitative agreement with observations (see Fig. \ref{f:hardness}). 

%4. The importance of the reflected component in the observed hardness 
%saturation in high-luminosity pulsars lend credence to the interpretation of the
%negative correlation of CRSF with X-ray luminosity in the reflection model
%%by Poutanen et al. (2013). No CRSF energy change is expected
%above the luminosity corresponding to the full illumination of the neutron
%star hemisphere. In the context of the calculations in the present paper,
%this corresponds to $\sim (5-8)\times 10^{17}$~g s$^{-1}$, where the
%spectral saturation is observed. However, change in the neutron star
%parameters and column geometry can shift this figures to higher values.

4. In the frame of this model, the saturation of
the spectral hardness in the case of a hollow cylinder geometry of the
accretion column can be achieved at much higher accretion rates (roughly,
scaled with the relative thickness of the column wall, $r_0/b$),
 because the characteristic height of the column in this
case is correspondingly smaller than that of the filled column. 

5. Variations in the pulse profile form and polarization
properties during the transition from the Coulomb-braking low-luminosity
regime to the radiation-braking high-luminosity state should occur 
with changing mass accretion rate (X-ray luminosity) in accreting X-ray pulsars. 
The profile change has already been reported during the low-states in Vela X-1
\citep{2011A&A...529A..52D}.
%(Doroshenko et al. 2011). 

We conclude that X-ray continuum observations in transient X-ray pulsars 
can be used as additional diagnostic of different accretion regimes 
and column geometry near the surface of highly magnetized neutron stars.

\section{Acknowledgements}

MG, NSh, VL and acknowledge the Russian Science Foundation grant 14-12-00146. The work of KP and DK is supported by RFBR/DFG grant KL 2734/2-1 (RFBR-NNIO 14-02-91345).

\bibliographystyle{mn2e}
%\expandafter\ifx\csname natexlab\endcsname\relax\def\natexlab#1{#1}\fi
\bibliography{column}

\begin{thebibliography}{42}
\expandafter\ifx\csname natexlab\endcsname\relax\def\natexlab#1{#1}\fi

\bibitem[{{Arons}(1992)}]{1992ApJ...388..561A}
{Arons} J., 1992, \apj, 388, 561

\bibitem[{{Basko} \& {Sunyaev}(1975)}]{1975A&A....42..311B}
{Basko} M.~M., {Sunyaev} R.~A., 1975, \aap, 42, 311

\bibitem[{{Basko} \& {Sunyaev}(1976)}]{BaskoSunyaev76}
{Basko} M.~M., {Sunyaev} R.~A., 1976, \mnras, 175, 395

\bibitem[{{Becker} {et~al}\mbox{.}(2012){Becker}, {Klochkov}, {Sch{\"o}nherr},
  {Nishimura}, {Ferrigno}, {Caballero}, {Kretschmar}, {Wolff}, {Wilms}, \&
  {Staubert}}]{Becker_ea12}
{Becker} P.~A. {et~al.}, 2012, \aap, 544, A123

\bibitem[{{Boldin}, {Tsygankov} \& {Lutovinov}(2013){Boldin}, {Tsygankov}, \&
  {Lutovinov}}]{2013AstL...39..375B}
{Boldin} P.~A., {Tsygankov} S.~S., {Lutovinov} A.~A., 2013, Astronomy Letters,
  39, 375

\bibitem[{{Caballero} \& {Wilms}(2012)}]{2012MmSAI..83..230C}
{Caballero} I., {Wilms} J., 2012, \memsai, 83, 230

\bibitem[{{Davidson}(1973)}]{1973NPhS..246....1D}
{Davidson} K., 1973, Nature Physical Science, 246, 1

\bibitem[{{Doroshenko}, {Santangelo} \& {Suleimanov}(2011){Doroshenko},
  {Santangelo}, \& {Suleimanov}}]{2011A&A...529A..52D}
{Doroshenko} V., {Santangelo} A., {Suleimanov} V., 2011, \aap, 529, A52

\bibitem[{{F{\"u}rst} {et~al}\mbox{.}(2014){F{\"u}rst}, {Pottschmidt}, {Wilms},
  {Tomsick}, {Bachetti}, {Boggs}, {Christensen}, {Craig}, {Grefenstette},
  {Hailey}, {Harrison}, {Madsen}, {Miller}, {Stern}, {Walton}, \&
  {Zhang}}]{2014ApJ...780..133F}
{F{\"u}rst} F. {et~al.}, 2014, \apj, 780, 133

\bibitem[{{Giacconi} {et~al}\mbox{.}(1971){Giacconi}, {Gursky}, {Kellogg},
  {Schreier}, \& {Tananbaum}}]{1971ApJ...167L..67G}
{Giacconi} R., {Gursky} H., {Kellogg} E., {Schreier} E., {Tananbaum} H., 1971,
  \apjl, 167, L67

\bibitem[{{Ho} \& {Lai}(2001)}]{2001MNRAS.327.1081H}
{Ho} W.~C.~G., {Lai} D., 2001, \mnras, 327, 1081

\bibitem[{{Klein} {et~al}\mbox{.}(1996){Klein}, {Arons}, {Jernigan}, \&
  {Hsu}}]{1996ApJ...457L..85K}
{Klein} R.~I., {Arons} J., {Jernigan} G., {Hsu} J.~J.-L., 1996, \apjl, 457, L85

\bibitem[{{Klochkov} {et~al}\mbox{.}(2012){Klochkov}, {Doroshenko},
  {Santangelo}, {Staubert}, {Ferrigno}, {Kretschmar}, {Caballero}, {Wilms},
  {Kreykenbohm}, {Pottschmidt}, {Rothschild}, {Wilson-Hodge}, \&
  {P{\"u}hlhofer}}]{2012A&A...542L..28K}
{Klochkov} D. {et~al.}, 2012, \aap, 542, L28

\bibitem[{{Klochkov} {et~al}\mbox{.}(2011){Klochkov}, {Staubert}, {Santangelo},
  {Rothschild}, \& {Ferrigno}}]{Klochkov:etal:11}
{Klochkov} D., {Staubert} R., {Santangelo} A., {Rothschild} R.~E., {Ferrigno}
  C., 2011, \aap, 532, A126

\bibitem[{{K{\"u}hnel} {et~al}\mbox{.}(2014){K{\"u}hnel}, {M{\"u}ller},
  {Kreykenbohm}, {F{\"u}rst}, {Pottschmidt}, {Rothschild}, {Caballero},
  {Grinberg}, {Sch{\"o}nherr}, {Shrader}, {Klochkov}, {Staubert}, {Ferrigno},
  {Torrej{\'o}n}, {Mart{\'{\i}}nez-N{\'u}{\~n}ez}, \&
  {Wilms}}]{2014EPJWC..6406003K}
{K{\"u}hnel} M. {et~al.}, 2014, in European Physical Journal Web of
  Conferences, Vol.~64, European Physical Journal Web of Conferences, p. 6003

\bibitem[{{Lutovinov} {et~al}\mbox{.}(2015){Lutovinov}, {Tsygankov},
  {Suleimanov}, {Mushtukov}, {Doroshenko}, {Nagirner}, \&
  {Poutanen}}]{2015MNRAS.448.2175L}
{Lutovinov} A.~A., {Tsygankov} S.~S., {Suleimanov} V.~F., {Mushtukov} A.~A.,
  {Doroshenko} V., {Nagirner} D.~I., {Poutanen} J., 2015, \mnras, 448, 2175

\bibitem[{{Lyubarskii}(1986)}]{1986Ap.....25..577L}
{Lyubarskii} Y.~{\'E}., 1986, Astrophysics, 25, 577

\bibitem[{{Lyubarskii} \& {Syunyaev}(1988)}]{1988SvAL...14..390L}
{Lyubarskii} Y.~E., {Syunyaev} R.~A., 1988, Soviet Astronomy Letters, 14, 390

\bibitem[{{McBride} {et~al}\mbox{.}(2006){McBride}, {Wilms}, {Coe},
  {Kreykenbohm}, {Rothschild}, {Coburn}, {Galache}, {Kretschmar}, {Edge}, \&
  {Staubert}}]{2006A&A...451..267M}
{McBride} V.~A. {et~al.}, 2006, \aap, 451, 267

\bibitem[{{Mihara}, {Makishima} \& {Nagase}(2004){Mihara}, {Makishima}, \&
  {Nagase}}]{2004ApJ...610..390M}
{Mihara} T., {Makishima} K., {Nagase} F., 2004, \apj, 610, 390

\bibitem[{{M{\"u}ller} {et~al}\mbox{.}(2013){M{\"u}ller}, {Ferrigno},
  {K{\"u}hnel}, {Sch{\"o}nherr}, {Becker}, {Wolff}, {Hertel}, {Schwarm},
  {Grinberg}, {Obst}, {Caballero}, {Pottschmidt}, {F{\"u}rst}, {Kreykenbohm},
  {Rothschild}, {Hemphill}, {N{\'u}{\~n}ez}, {Torrej{\'o}n}, {Klochkov},
  {Staubert}, \& {Wilms}}]{2013A&A...551A...6M}
{M{\"u}ller} S. {et~al.}, 2013, \aap, 551, A6

\bibitem[{{Mushtukov} {et~al}\mbox{.}(2015){Mushtukov}, {Suleimanov},
  {Tsygankov}, \& {Poutanen}}]{2015MNRAS.447.1847M}
{Mushtukov} A.~A., {Suleimanov} V.~F., {Tsygankov} S.~S., {Poutanen} J., 2015,
  \mnras, 447, 1847

\bibitem[{{Nagel}(1980)}]{1980ApJ...236..904N}
{Nagel} W., 1980, \apj, 236, 904

\bibitem[{{Negueruela} \& {Okazaki}(2001)}]{Negueruela:Okazaki:01}
{Negueruela} I., {Okazaki} A.~T., 2001, \aap, 369, 108

\bibitem[{{Negueruela} {et~al}\mbox{.}(1999){Negueruela}, {Roche}, {Fabregat},
  \& {Coe}}]{Negueruela:etal:99}
{Negueruela} I., {Roche} P., {Fabregat} J., {Coe} M.~J., 1999, \mnras, 307, 695

\bibitem[{{Nelson}, {Salpeter} \& {Wasserman}(1993){Nelson}, {Salpeter}, \&
  {Wasserman}}]{1993ApJ...418..874N}
{Nelson} R.~W., {Salpeter} E.~E., {Wasserman} I., 1993, \apj, 418, 874

\bibitem[{{Nishimura}(2014)}]{2014ApJ...781...30N}
{Nishimura} O., 2014, \apj, 781, 30

\bibitem[{{Parkes}, {Murdin} \& {Mason}(1980){Parkes}, {Murdin}, \&
  {Mason}}]{Parkes:etal:80}
{Parkes} G.~E., {Murdin} P.~G., {Mason} K.~O., 1980, \mnras, 190, 537

\bibitem[{{Poutanen} {et~al}\mbox{.}(2013){Poutanen}, {Mushtukov},
  {Suleimanov}, {Tsygankov}, {Nagirner}, {Doroshenko}, \&
  {Lutovinov}}]{2013ApJ...777..115P}
{Poutanen} J., {Mushtukov} A.~A., {Suleimanov} V.~F., {Tsygankov} S.~S.,
  {Nagirner} D.~I., {Doroshenko} V., {Lutovinov} A.~A., 2013, \apj, 777, 115

\bibitem[{{Reig} \& {Nespoli}(2013)}]{Reig:Nespoli:13}
{Reig} P., {Nespoli} E., 2013, \aap, 551, A1

\bibitem[{{Revnivtsev} \& {Mereghetti}(2014)}]{2014SSRv..tmp...58R}
{Revnivtsev} M., {Mereghetti} S., 2014, \ssr

\bibitem[{{Staubert} {et~al}\mbox{.}(2007){Staubert}, {Shakura}, {Postnov},
  {Wilms}, {Rothschild}, {Coburn}, {Rodina}, \&
  {Klochkov}}]{2007A&A...465L..25S}
{Staubert} R., {Shakura} N.~I., {Postnov} K., {Wilms} J., {Rothschild} R.~E.,
  {Coburn} W., {Rodina} L., {Klochkov} D., 2007, \aap, 465, L25

\bibitem[{{Steele} {et~al}\mbox{.}(1998){Steele}, {Negueruela}, {Coe}, \&
  {Roche}}]{1998MNRAS.297L...5S}
{Steele} I.~A., {Negueruela} I., {Coe} M.~J., {Roche} P., 1998, \mnras, 297, L5

\bibitem[{{Tsygankov} {et~al}\mbox{.}(2006){Tsygankov}, {Lutovinov},
  {Churazov}, \& {Sunyaev}}]{2006MNRAS.371...19T}
{Tsygankov} S.~S., {Lutovinov} A.~A., {Churazov} E.~M., {Sunyaev} R.~A., 2006,
  \mnras, 371, 19

\bibitem[{{Tsygankov} {et~al}\mbox{.}(2007){Tsygankov}, {Lutovinov},
  {Churazov}, \& {Sunyaev}}]{2007AstL...33..368T}
{Tsygankov} S.~S., {Lutovinov} A.~A., {Churazov} E.~M., {Sunyaev} R.~A., 2007,
  Astronomy Letters, 33, 368

\bibitem[{{Tsygankov}, {Lutovinov} \& {Serber}(2010){Tsygankov}, {Lutovinov},
  \& {Serber}}]{2010MNRAS.401.1628T}
{Tsygankov} S.~S., {Lutovinov} A.~A., {Serber} A.~V., 2010, \mnras, 401, 1628

\bibitem[{{Ventura}(1979)}]{1979PhRvD..19.1684V}
{Ventura} J., 1979, \prd, 19, 1684

\bibitem[{{Virtamo} \& {Jauho}(1975)}]{1975NCimB..26..537V}
{Virtamo} J., {Jauho} P., 1975, Nuovo Cimento B Serie, 26, 537

\bibitem[{{Wang} \& {Frank}(1981)}]{1981A&A....93..255W}
{Wang} Y.-M., {Frank} J., 1981, \aap, 93, 255

\bibitem[{{Wilson} {et~al}\mbox{.}(2002){Wilson}, {Finger}, {Coe}, {Laycock},
  \& {Fabregat}}]{Wilson:etal:02}
{Wilson} C.~A., {Finger} M.~H., {Coe} M.~J., {Laycock} S., {Fabregat} J., 2002,
  \apj, 570, 287

\bibitem[{{Yamamoto} {et~al}\mbox{.}(2011){Yamamoto}, {Sugizaki}, {Mihara},
  {Nakajima}, {Yamaoka}, {Matsuoka}, {Morii}, \&
  {Makishima}}]{2011PASJ...63S.751Y}
{Yamamoto} T., {Sugizaki} M., {Mihara} T., {Nakajima} M., {Yamaoka} K.,
  {Matsuoka} M., {Morii} M., {Makishima} K., 2011, \pasj, 63, 751

\bibitem[{{Zel'dovich} \& {Shakura}(1969)}]{1969SvA....13..175Z}
{Zel'dovich} Y.~B., {Shakura} N.~I., 1969, \sovast, 13, 175

\end{thebibliography}

\appendix
\section{X-ray albedo from neutron star atmosphere}
\label{s:appendix}

The X-ray albedo (the fraction of the reflected to incident flux)
from the neutron star atmosphere in strong magnetic field
in the general case depends on
the angle between the incident photon and the magnetic field and the
photon polarization. Not solving exactly the problem, we calculate
single-scattering albedo for a plane-parallel atmosphere
$\lambda^j(\nu)$ ($j=1,2$ for extraordinary and ordinary
photon polarizations, respectively):
\beq{}
\lambda^j(\nu)=\frac{\kappa_{sc}^j}{\kappa_{sc}^j+\kappa_{abs}^j},
\eeq
where $\kappa_{sc}^j$ and $\kappa_{abs}^j$ are
scattering and absorption coefficients in the magnetic field,
respectively.

We calculate the coefficients following \cite{1979PhRvD..19.1684V, 1980ApJ...236..904N, 2001MNRAS.327.1081H}.
The electron and ion scattering coefficients for mode $j$ reads:
\beq{k_esc}
\kappa_{sc,\,e}^j=n_e\sigma_T\sum\limits_{\alpha=-1}^1\frac{\omega^2}{(\omega+\alpha\omega_{c,\,e})^2}|e_\alpha^j|^2 %\frac{\omega^2}{(\omega-\omega_{c,e})^2}|e_-^j|^2 + |e_z^j|^2 ,
\eeq
\beq{k_isc}
\kappa_{sc,\,i}^j=\left(\frac{Z^2m_e}{Am_p}\right)^2n_i\sigma_T\sum\limits_{\alpha=-1}^1\frac{\omega^2}{(\omega+\alpha\omega_{c,\,i})^2}|e_\alpha^j|^2 %+ \frac{\omega^2}{(\omega-\omega_{c,i})^2}|e_-^j|^2 + |e_z^j|^2,
\eeq
where $\sigma_T$ is the Thomson scattering cross-section, $m_e$ and $m_p$ is the electron and proton masses,  $n_e$ and $n_i$ is electron and ion concentrations,
$\omega_{c,\,e}$ and $\omega_{c,\,i}$ is the electron and ion cyclotron angular frequency, $e_{-1}^j$, $e_1^j$, are $e_0^j$ are the components of the photon polarization vector $\mathbf{e}^j$ for mode $j$ (see subsection 2.5 and Appendix A in \citep{2001MNRAS.327.1081H}). The absorption coefficients are
\beq{k_eabs}
\kappa_{abs,\,e}^j=\kappa_0\sum\limits_{\alpha=-1}^1\frac{\omega^2}{(\omega+\alpha\omega_{c,\,e})^2}|e_\alpha^j|^2 g_\alpha %+\frac{\omega^2}{(\omega-\omega_{c,e})^2}|e_-^j|^2 g_\bot+ |e_z^j|^2 g_\parallel \right],
\eeq
\beq{k_iabs}
\kappa_{abs,\,i}^j=\left(\frac{Z^2m_e}{Am_p}\right)^2\frac{\kappa_0}{Z^3}\sum\limits_{\alpha=-1}^1
\frac{\omega^2}{(\omega+\alpha\omega_{c,\,i})^2}|e_\alpha^j|^2 g_\alpha
%+ \frac{\omega^2}{(\omega-\omega_{c,i})^2}|e_-^j|^2 g_\bot+ |e_z^j|^2 g_\parallel \right],
\eeq
where
\beq{kappa_0}
\kappa_0=4\pi^2 Z^2 \alpha^3 \left(\frac{\hbar c}{m_e}\right)^2 \left(\frac{2m_e}{\pi k T}\right)^{1/2}\frac{n_en_i}{\omega^3}
\left(1-e^{-\frac{\hbar \omega}{kT}}\right)
\eeq
is the free-free absorption coefficient  in the absence of a magnetic field, $g_{\pm 1}$ and $g_0$ are the modified Gaunt-factors \citep{1980ApJ...236..904N},
\beq{}
g_{\pm 1}=\int\limits_{-\infty}^\infty \exp\left(-\frac{\hbar\omega}{kT}\sinh^2 x\right)C_1\left(\frac{\omega}{\omega_c}e^{2x}\right)dx,
\eeq
\beq{}
g_0=\int\limits_{-\infty}^\infty \exp\left(-\frac{\hbar\omega}{kT}\sinh^2 x\right)2\frac{\omega}{\omega_c}e^{2x}C_0\left(\frac{\omega}{\omega_c}e^{2x}\right)dx.
\eeq
Here $\omega_c$ is electron (ion) cyclotron angular frequency.
Functions $C_1$ and $C_0$ are calculated in \cite{1975NCimB..26..537V}.
The total coefficients are:
\beq{}
\kappa_{sc}^j=\kappa_{sc,\,e}^j+\kappa_{sc,\,i}^j,
\eeq
\beq{}
\kappa_{abs}^j=\kappa_{abs,\,e}^j+\kappa_{abs,\,i}^j.
\eeq

The coefficients  $\kappa_{sc}$ and $\kappa_{abs}$ depend on many factors.
They are determined by the characteristic of reflecting atmosphere, including the chemical composition, ion number density and temperature, as well as by
the photon incident angle to the magnetic field and the magnetic field strength.
The last two parameters turn out to be weakly changing the X-ray albedo
$\lambda^j(\nu)$  in the interesting range of parameters (see Fig. \ref{f:albedo}).
In the calculations we have assumed purely hydrogen atmosphere with ion charge
$Z=1$, the mass number $A=1$ and the ion number density
$n_i=5\times10^{25}$~cm$^{-3}$ (it can be higher
in deeper layers \citep{2001MNRAS.327.1081H}). The electron
temperature is fixed at $T=3$~keV, which corresponds to
the expected Compton temperature during reflection of X-ray emission form
the column.

The single-scattering X-ray albedo as a function of frequency
for ordinary ($j=2$) and extraordinary ($j=1$) photon polarizations
is shown in Fig. \ref{f:albedo}.

\end{document}